\documentclass[%
 reprint,
 amsmath,amssymb,
 aps,
]{revtex4-2}

\usepackage{amsmath}
\usepackage{physics}
\usepackage{graphicx}
\usepackage{dcolumn}
\usepackage{bm}

\begin{document}
\title{Operating Principles of Peristaltic Pumping through a Dense Array of Valves}

\author{Aaron Winn}
\email{winna@sas.upenn.edu}
\author{Eleni Katifori}

\affiliation{%
 Department of Physics and Astronomy, University of Pennsylvania, Philadelphia, Pennsylvania 19104, USA
}%

\date{\today}
\begin{abstract}
Immersed nonlinear elements are prevalent in biological systems that require a preferential flow direction, such as the venous and the lymphatic system. We investigate here a certain class of models where the fluid is driven by peristaltic pumping and the nonlinear elements are ideal valves that completely suppress backflow. This highly nonlinear system produces discontinuous solutions that are difficult to study. We show that as the density of valves increases, the pressure and flow are well-approximated by a continuum of valves which can be analytically treated, and we demonstrate through numeric simulation that the approximation works well even for intermediate valve densities. We find that the induced flow is linear in the peristaltic amplitude for small peristaltic forces and, in the case of sinusoidal peristalsis, is independent of pumping direction. Despite the continuum approximation used, the physical valve density is accounted for by modifying the resistance of the fluid appropriately. The suppression of backflow causes a net benefit in adding valves when the valve density is low, but once the density is high enough, valves predominately suppress forward flow, suggesting there is an optimum number of valves per wavelength. The continuum model for peristaltic pumping through an array of valves presented in this work can eventually provide insights about the design and operating principles of complex flow networks with a broad class of nonlinear elements. 
\end{abstract}
\maketitle

\section{Introduction}
\label{sec:intro}

\begin{figure*}
    \centering
    \includegraphics[width=.65\textwidth]{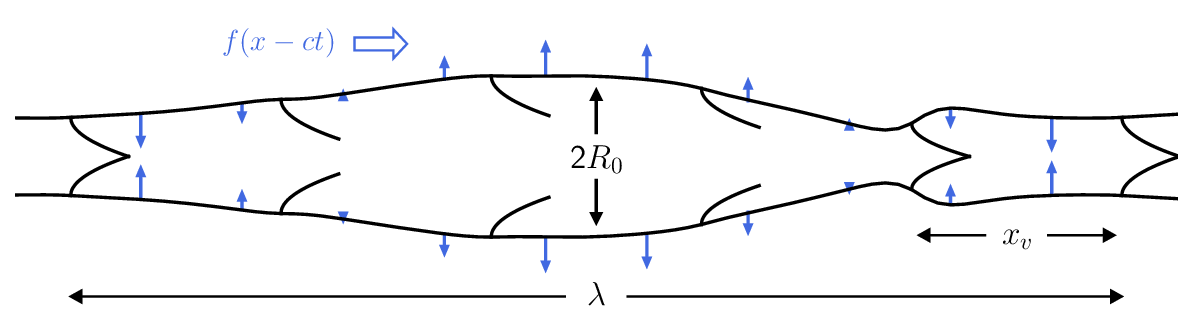}
    \caption{Model for peristaltic pumping with a dense array of valves satisfying assumption \eqref{eq:approx}. The imposed peristaltic force $f$ has wavelength $\lambda$ and speed $c$. The strength of the force at each point on the boundary is proportional to the length of the arrows. The vessel has rest radius $R_0$ and intervalve spacing $x_v$. }
    \label{fig:model}
\end{figure*}

Peristalsis occurs when external radial forces propagate along a fluid-filled tube, inducing fluid motion. In the human body alone, peristaltic waves drive fluid transport in the esophagus \citep{Brasseur_1987}, the ureter \citep{Carew_Pedley_1997}, the lymphatic system \citep{Moore_Bertram_2018}, and the perivascular spaces of the brain \citep{carr_thomas_liu_shang_2021,Mestre2018}. The success of modeling peristaltic pumping arises from its simplicity: When only peristaltic forces drive flow and the forces take the form of a wave propagating in an infinitely long tube, the Navier Stokes equations describe steady flow in the co-moving wave frame. The problem of peristalsis at low Reynolds number was first studied perturbatively in powers of a small-amplitude parameter \citep{Burns_Parkes_1967} and later extended to the case of arbitrary amplitudes, but under the assumption of long wavelength \citep{Shapiro_Jaffrin_Weinberg_1969}. Historically, the term ``long-wavelength peristalsis" has been used to refer to a regime where the wavelength is large compared to the unperturbed radius of the compliant tube $R_0$. When nonlinear elements are scattered throughout the tube, one has an additional length scale arising from the characteristic spacing between these elements, $x_v$ in figure \ref{fig:model}. These nonlinear elements are important for determining the pressure-flow relationship, but they introduce complexities that render an analytical treatment difficult. In this paper, we will consider the case when these nonlinear elements are ideal valves that completely prevent backflow.

The combination of peristalsis and valves at low Reynolds number is particularly relevant for studying biological fluid networks with a nonlinear pressure-flow relationship. For example, synchronous peristalsis in a finite tube capped with valves at both ends was used to model a bat wing venule \citep{Farina_Fusi_Fasano_Ceretani_Rosso_2016}. Our model closely resembles pumping in the collecting lymphatic vessels where intrinsic and extrinsic pumping mechanisms transport lymph through units separated by valves called lymphangions \citep{Moore_Bertram_2018, Margaris_Black_2012}. The peristaltic response in the lymphatic system is fundamentally different from that in the esophagus or the ureter in that the direction of fluid motion is fixed by the valve orientation, not the direction of peristaltic wave propagation. The fluid is transported in the valve direction even when the peristaltic wave travels in the opposite direction \citep{McHale_Meharg_1992,  Zawieja_Davis_Schuster_Hinds_Granger_1993}. Lymphatic-inspired peristaltic pumping with valves has been studied numerically using a lattice Boltzmann model for the lymph and a lattice spring model for the mechanics \citep{Ballard_Wolf_Nepiyushchikh_Dixon_Alexeev_2018, Wolf_Dixon_Alexeev_2021, Wolf_Poorghani_Dixon_Alexeev_2023}. The aspect ratio \citep{Ballard_Wolf_Nepiyushchikh_Dixon_Alexeev_2018}, bending stiffness \citep{Wolf_Dixon_Alexeev_2021}, and spacing \citep{Wolf_Poorghani_Dixon_Alexeev_2023} of the valves all play a role in enhancing the net flow and energetic efficiency of the lymphatic system. The valves must be designed in such a way that backflow is prevented when pressure is unfavorable while keeping the resistance to forward flow minimal. 

Existing numerical models elucidate operating principles of the lymphatic system, but due to the nonlinearity of the valves, no analytical treatment of peristalsis with many valves has been attempted. In this work, we will assume that the peristaltic wavelength $\lambda$ is much longer than the characteristic valve spacing $x_v$: \begin{equation}
    R_0 \ll x_v \ll \lambda .\label{eq:approx}
\end{equation}
The first inequality allows us to neglect complicated behavior near the valves and apply the lubrication approximation. The valves are spaced far enough apart that the velocity profile remains parabolic throughout most of the channel with no slip at the tube walls, and the resistances of the valves can be added in series. The second inequality suggests that the valves are dense enough that we can write approximate expressions for the flow through many closed or open valves. See figure \ref{fig:model} for an example geometry that satisfies approximation \eqref{eq:approx}.

While focus will be given to this lymphatics-inspired model, the technique demonstrated in this paper could be used for a variety of problems containing a dense array of nonlinear elements satisfying \eqref{eq:approx}, and could be of potential interest for engineering applications that incorporate artificial valves (see e.g. \cite{Park2018,Brandenbourger2020}). The key observation will be that when \eqref{eq:approx} is satisfied, the precise placement of valves becomes unimportant, and the flow is well-approximated by treating the entire medium as a fluid with nonlinear properties inherited from the valves. This effective nonlinear fluid will be referred to as the \textit{valve continuum}. While a finite number of valves will break the translation symmetry required to study steady flow in the co-moving wave frame, this symmetry is restored in the valve continuum, allowing us to make analytical progress into this highly nonlinear problem. The valve continuum has peculiar properties which we will analyze throughout the paper. Perhaps most interesting is the property that a backward-propagating peristaltic wave can induce flow in the forward direction (the valve's preferred direction) of comparable magnitude to the flow induced by a forward-propagating peristaltic wave, elucidating the peculiar operating principle observed in the lymphatic system. 

The paper is outlined as follows. In section \ref{sec:peristalsis}, the fluid and solid equations governing force-imposed peristaltic pumping in an elastic tube are reviewed, along with the choice of nondimensionalization. In section \ref{sec:valves}, the equations for discrete ideal valves are introduced. It is then demonstrated how to approximately describe the fluid confined to regions of many closed valves in \ref{sec:closed_valves}, many open valves in \ref{sec:open_valves}, and appropriate matching conditions in \ref{sec:matching}. From these considerations, one arrives at a model for the valve continuum. Throughout sections \ref{sec:valve_continuumFW} and \ref{sec:valve_continuumBW}, solutions to the valve continuum model are studied for the cases of forward-propagating and backward-propagating peristaltic waves, respectively. Explicit solutions are found and plotted for the special case of sinusoidal peristaltic waves. The role of the open valve resistance in setting the optimum valve density is discussed in section \ref{sec:density}. Finally, discussion on how this model relates to the lymphatic system and other applications is given in section \ref{sec:discussion}. Additional mathematical details and a table of parameters are given in the appendix. 

\section{Force-Imposed Peristalsis at Low Reynolds Number} \label{sec:peristalsis}
There are two methods of mathematically modeling peristalsis on a cylindrical pipe. The most commonly used model assumes that the radius varies in time according to some prescribed function in the form of a wave $R(x-ct)$, where $c$ is the wave speed. This induces fluid motion in the tube, and the pressure and flow can be easily calculated \citep{Shapiro_Jaffrin_Weinberg_1969, Burns_Parkes_1967}. This method is appropriate for modeling the response from a peristaltic pump where the radius is fixed by the size of the rollers, but in the biological setting, it is more accurate to measure the fluid response from a force per area propagating along the pipe. This captures the fluid-structure interaction at the walls of the vessel. The goal of the paper will be to generalize the results of force-imposed peristalsis \citep{Takagi_Balmforth_2011, Elbaz_Gat_2014, Carew_Pedley_1997} to the case with valves. In order to isolate the effects of peristalsis, the mean pressure drop per wavelength will be assumed zero throughout the paper.

\subsection{Dimensional Formulation}
We will concern ourselves only with an incompressible fluid at low Reynolds number under the lubrication approximation. Since the radial velocity is always small, the pressure is only a function of the axial coordinate $x$, and the velocity profile is assumed to remain parabolic. Thus, it is sufficient to work only in terms of the flow $Q(x,t)$, since the axial velocity $u_x(x,r,t)$ can be recovered by using the following relations: 
\begin{equation}
    Q(x,t) \equiv \int_0^{R(x,t)} u_x(x,r,t) 2\pi r dr ,
\end{equation}
\begin{equation}
    u_x (x,r,t) = \frac{2Q(x,t)}{\pi R(x,t)^2} \left(1 - \frac{r^2}{R(x,t)^2} \right).
\end{equation}
Here, $R(x,t)$ is the radius of the tube, and $r$ is the distance from the midline of the vessel. Under our approximations, the equations governing mass continuity and momentum conservation reduce to 
\begin{equation}
    \frac{\partial Q}{\partial x} + \pi \frac{\partial R^2}{\partial t} = 0, \label{eq:cont}
\end{equation}
\begin{equation}
    \frac{\partial P}{\partial x} + \frac{8 \mu}{\pi R^4}  Q = 0, \label{eq:mom}
\end{equation}
where $\mu$ is the dynamic viscosity. Finally, the pressure and radius are coupled via a linear elasticity equation \citep{Timoshenko_1959, Takagi_Balmforth_2011}:
\begin{equation}
    P-P_a f(x,t) = \frac{Eh}{(1-\nu^2)R_0}  \Big(\frac{R}{R_0}-1\Big), \label{eq:force_balance}
\end{equation}
where $E$ is the Young's modulus, $h$ is the thickness of the tube, $\nu$ is the Poisson's ratio, and $R_0$ is the rest radius of the tube, which in this work is assumed to be independent of $x$. A generalization of \eqref{eq:force_balance} was considered by \cite{Macdonald_Arkill_Tabor_McHale_Winlove_2008} to represent the elastic response of lymphangions; the form is still a linear relationship between pressure and radial deformation, but with a different coefficient. $P_a f$ is the prescribed peristaltic force with characteristic amplitude $P_a$ and functional form $f$ whose mean value is zero. Throughout the paper, most of our numerical results for a forward-propagating wave will use $f(x,t) = \cos(2\pi (x-ct))$, and our results for a backward-propagating wave will use $f(x,t) = -\cos(2\pi (x+ct))$. 

Consequences of adding a small bending term to \eqref{eq:force_balance} will be discussed in the appendix.

\subsection{Dimensionless Formulation}
We will now work with convenient dimensionless quantities:
\begin{widetext}
\begin{equation}
    \Bar{R} \equiv \frac{R}{R_0}, \hspace{1cm} \Bar{Q} \equiv \frac{Q}{c\pi R_0^2}, \hspace{1cm} \Bar{P} \equiv \frac{P}{(c\pi R_0^2)(8\mu \lambda  /\pi R_0^4)}, \label{eq:PQR_nondim}
\end{equation}
\begin{equation}
    \Bar{x} \equiv \frac{x}{\lambda} ,\hspace{1cm} \Bar{t} \equiv \frac{t}{T}, \label{eq:xt_nondim}
\end{equation}
\begin{equation}
     \kappa \equiv \frac{Eh/(1-\nu^2)R_0}{(c\pi R_0^2)(8\mu \lambda  /\pi R_0^4)}, \hspace{.5cm} \eta_P \equiv \frac{P_a}{(c\pi R_0^2)(8\mu \lambda  /\pi R_0^4)} , \hspace{.5cm} \eta_R \equiv  \frac{\eta_P}{\kappa} = \frac{P_a}{Eh/(1-\nu^2)R_0} . \label{eq:parameters}
\end{equation}
\end{widetext}
The radius, flow, and pressure nondimensionalization is similar to that used in previous papers describing peristalsis, such as \cite{Shapiro_Jaffrin_Weinberg_1969, Provost_Schwarz_1994}.
Space is scaled by the peristaltic wavelength $\lambda$, and time is scaled by the peristaltic period $T$. $\kappa$ is the ratio of the stiffness to the characteristic pressure of peristalsis in a viscous tube, $\eta_P$ is the ratio of the applied peristaltic force to the characteristic pressure of peristalsis in a viscous tube, and $\eta_R$ (which can be constructed from the other two parameters) gives the characteristic radial deformation of a stiff vessel.

Using these dimensionless variables, our model for peristaltic pumping in an elastic tube becomes
\begin{equation}
    \frac{\partial \bar{Q}}{\partial \bar{x}} +   \frac{\partial \bar{R}^2}{\partial \bar{t}} = 0 ,\label{eq:continuity}
\end{equation}
\begin{equation}
    \bar{Q} = -\bar{R}^4 \frac{\partial \bar{P}}{\partial \bar{x}}  ,\label{eq:momentum}
\end{equation}
\begin{equation}
    \bar{P}=\kappa (\bar{R}-1 ) + \eta_P f, \hspace{.5cm} \Bar{R} = 1-\eta_R f + \frac{1}{\kappa} \Bar{P}\label{eq:PR}.
\end{equation}
When periodic boundary conditions are applied, the integral of \eqref{eq:continuity} implies that the total volume is conserved. By convention (or by an appropriate definition of $R_0$), we will enforce this dimensionless volume to remain one 
\begin{equation}
    \int_0^1 \Bar{R}^2 d\Bar{x} = 1. \label{eq:RSquaredConstraint}
\end{equation}
This simple form of the volume constraint was also imposed in \cite{Takagi_Balmforth_2011}. Keeping only linear terms in the radial deformation allows one to combine equations \eqref{eq:continuity}, \eqref{eq:momentum}, and \eqref{eq:PR} into a single driven heat equation:
\begin{equation}
    \frac{\partial \bar{P}}{\partial \bar{t}} - \frac{\kappa}{2} \frac{\partial^2 \bar{P}}{\partial \bar{x}^2} \approx \eta_P \frac{\partial f}{\partial \bar{t}} .\label{eq:diffusion}
\end{equation}
Although this equation is only true to linear order in $\eta_P$, it allows us to better understand the role of the parameter $\kappa$. One can think of $\kappa/2$ as a diffusion coefficient or $2/\kappa$ as an elastic relaxation time per peristaltic period. When $\kappa$ is large, the pressure diffuses to an equilibrium configuration of nearly uniform pressure. When $\kappa$ is small, the relaxation time is longer than the peristaltic period, and the pressure distribution closely resembles the external pressure applied to the tube. 

\section{Incorporating Valves into a Model for Peristalsis} \label{sec:valves}
A valve could be any nonlinear element that promotes flow in one direction more than the other. We will concern ourselves only with the extreme case of an ideal valve, which only allows flow in one direction proportional to the pressure drop across the valve. This is equivalent both to the diode representation of valves used in lumped models of the lymphatic system \citep{Margaris_Black_2012} and to the boundary conditions used by \cite{Farina_Fusi_Fasano_Ceretani_Rosso_2016} to study veinous valves. A more detailed valve model could incorporate the mechanical properties of the valves such as the bending and stretching stiffness \citep{Wolf_Dixon_Alexeev_2021}.  Superscripts are used to index a particular valve; specifically, $x_v^i$ is used to denote the position of valve $i$. For our ideal valves, the valve status (open or closed) is determined by the sign of the pressure drop across the valve, such that the valve closes when the pressure downstream exceeds the pressure upstream, and the valve opens when the pressure upstream exceeds the pressure downstream. Formally, if we let $\Delta P_v^i(t)$ denote the upstream pressure minus the downstream pressure, then the valves open and close according to the following equations:
\begin{align}
    &\Delta P_v^i (t) = 0 \hspace{.25cm} \text{and} \hspace{.25cm} \frac{\partial}{\partial t} \Delta P_v^i (t) < 0  \hspace{.25cm} \implies \hspace{.25cm} \text{valve closes} , \label{eq:closes}\\ 
    &\Delta P_v^i (t) = 0 \hspace{.25cm} \text{and} \hspace{.25cm} \frac{\partial}{\partial t} \Delta P_v^i (t) > 0  \hspace{.25cm} \implies \hspace{.25cm} \text{valve opens}.\label{eq:opens}
\end{align}
A closed valve has identically zero flow and a negative pressure drop. An open valve has a positive pressure drop related to the flow by Poiseuille's law. Here, we will not concern ourselves with the fluid dynamics inside the valve, but instead assume the pressure-flow relationship at valve $i$ is
\begin{equation}
     \mathcal{R}_v Q_v^i(t) = \left(\frac{R(x_v^i,t)}{R_0}\right)^4 \Delta P_v^i(t) \Theta( \Delta P_v^i(t) ),
\end{equation}
where $\mathcal{R}_v$ is the resistance of a fully open valve. Using the nondimensional functions from the previous section, 
\begin{equation}
    \bar{r}_v  \bar{Q}_v^i (\bar{t})= \bar{R}(\bar{x}_v^i, \bar{t})^4 \Delta \bar{P}_v^i (\bar{t}) \Theta  \big(\Delta \bar{P}_v^i (\bar{t}) \big) , \label{eq:valve}
\end{equation}
where $\Bar{r}_v$ is a dimensionless resistance parameter which compares the resistance of an open valve to that of a valveless tube of length $\lambda$:

\begin{equation}
    \Bar{r}_v \equiv \frac{\mathcal{R}_v}{8\mu \lambda/\pi R_0^4}. \label{eq:rv}
\end{equation}

\begin{figure*}
    \centering
    \includegraphics[width=.45\textwidth]{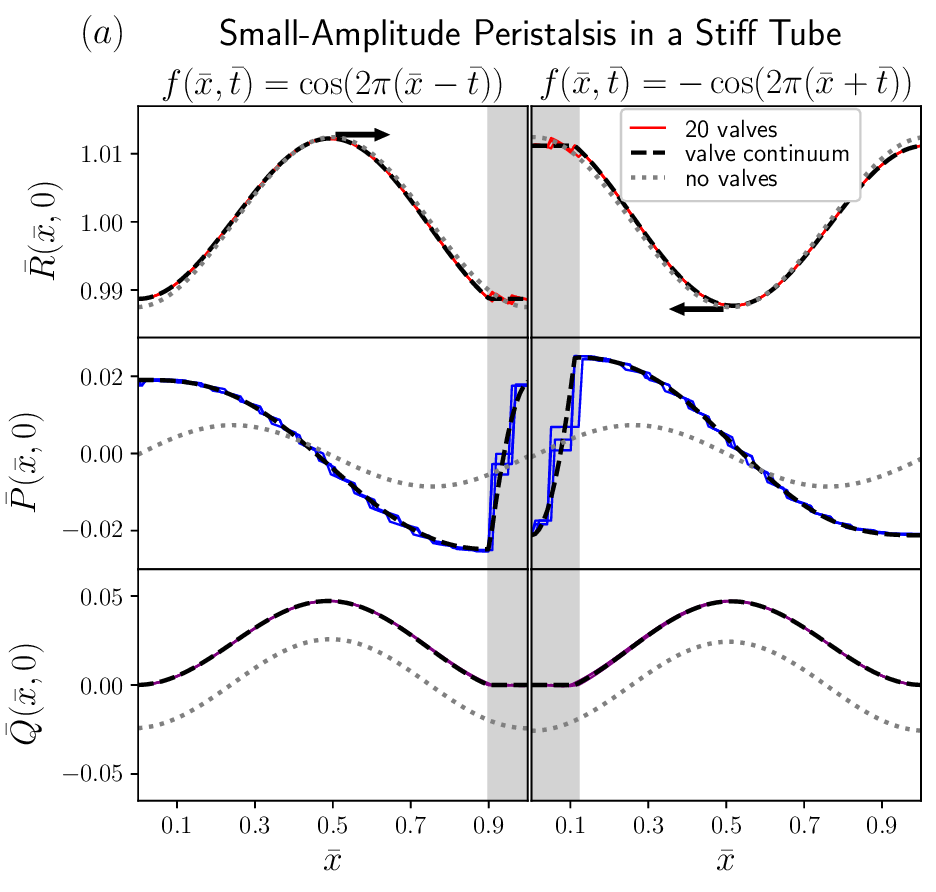}
    \includegraphics[width=.45\textwidth]{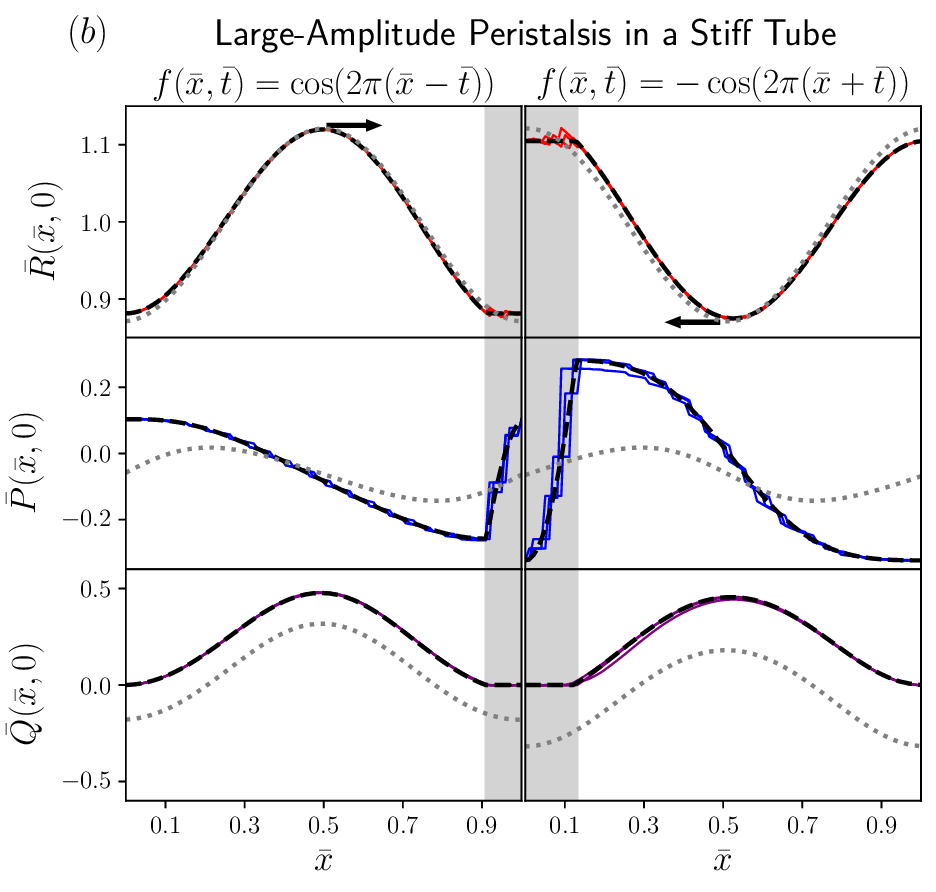}
    \includegraphics[width=.45\textwidth]{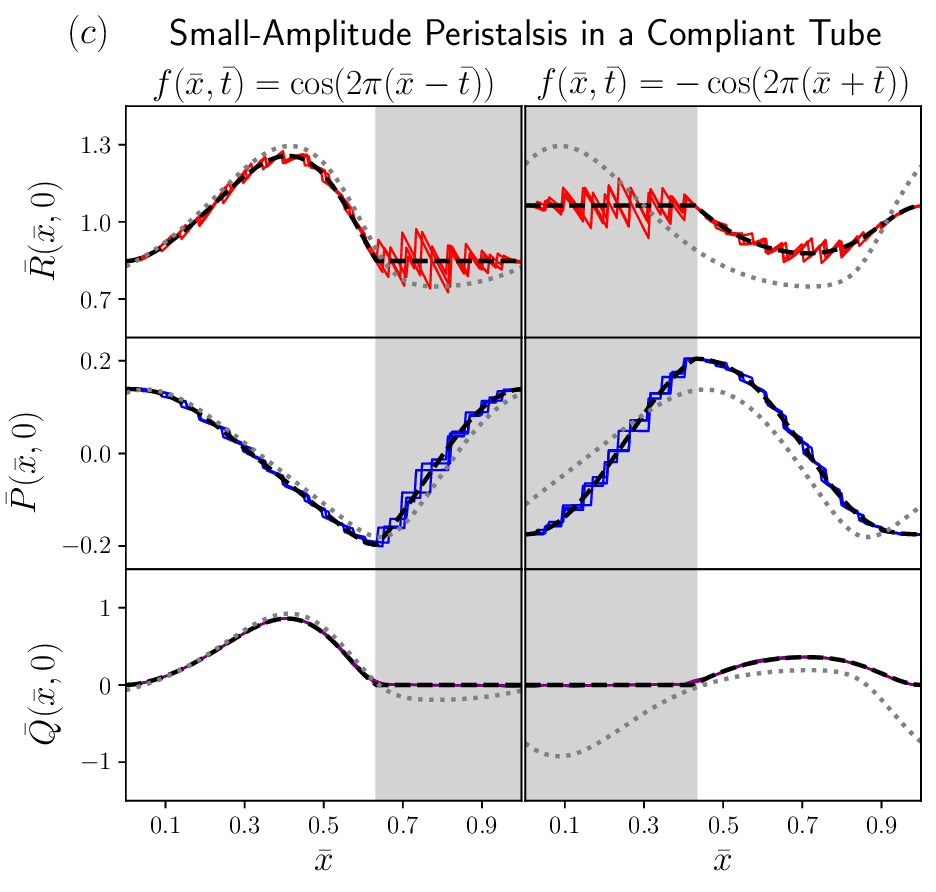}
    \includegraphics[width=.45\textwidth]{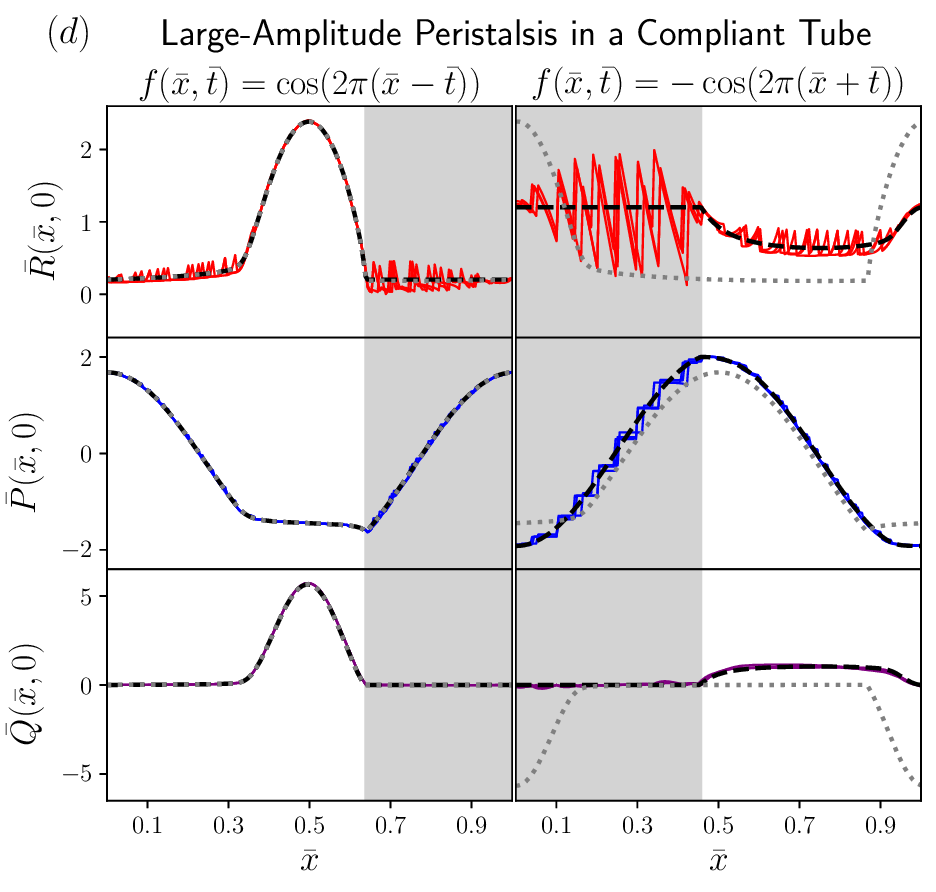}
    \caption{Demonstration of how the pressure and flow in a system of many randomly placed valves is approximated by the results in the valve continuum limit for various choices of the amplitude of peristalsis $\eta_P$ and the stiffness $\kappa$. The solid lines show numerical results from simulating a tube with $n_v=20$ valves per wavelength. Red solid line: normalized radius as a function of $\bar{x}$, the normalized location along the tube; blue solid line: normalized pressure; purple solid line: normalized flow. The dimensionless valve resistance was chosen to be $\bar{r}_v = .05$. The black dashed line shows the corresponding valve continuum prediction, and the gray dotted line shows the prediction for the valveless case. The valves in the shaded region are closed while the valves in the unshaded region are open. Parameters used for the discrete valve simulations are $(a)$ $\kappa = 16, \eta_P=.2$, $(b)$ $\kappa = 16, \eta_P=2$, $(c)$ $\kappa = .4, \eta_P=.2$, $(d)$ $\kappa = .4, \eta_P=2$. For the valve continuum and valveless cases, $\kappa$ and $\eta_P$ were divided by $1+\bar{r}_v n_v=2$, and $\bar{P}$ was multiplied by 2 as compared to the discrete valve case. }
    \label{fig:convergence}
\end{figure*}

In order to numerically solve equations \eqref{eq:continuity}, \eqref{eq:momentum}, \eqref{eq:PR}, and \eqref{eq:valve}, we discretize space with periodic boundary conditions and integrate in time using Scipy's solve\_ivp function initialized with the tube at rest. At each time step, the area $\Bar{R}^2$ is updated according to \eqref{eq:continuity}, the pressure is calculated from \eqref{eq:PR}, and the flow is calculated using \eqref{eq:momentum} and \eqref{eq:valve}. Note that edges without valves have a fluidic resistance equal to the step size $d \bar{x}$, edges with open valves have a fluidic resistance of $\Bar{r}_v+ d \bar{x}$, and edges with closed valves have an infinite fluidic resistance. The radius along an edge with a valve is multivalued, but numerically, the areas upstream and downstream are averaged when computing the factor of $\Bar{R}^4$ in \eqref{eq:valve}. Integration stops once the system has converged to its periodic steady state.

Solutions to these equations using $f(\bar{x},\bar{t}) = \pm \cos (2\pi (\bar{x}\mp \bar{t}))$ are displayed in figure \ref{fig:convergence}. Twenty valves were placed on the domain with mean valve separation $\epsilon \equiv x_v/\lambda=.05$ and standard deviation .01. The resistance of a single valve was chosen to be $\Bar{r}_v=.05$ such that the total valve resistance is $n_v \Bar{r}_v = 1$, where $n_v =\epsilon^{-1}$ is the number of valves per wavelength. For each choice of parameters, three solutions differing only by their random choice of valve placements were plotted on top of each other with solid lines. The details of the various regimes will be explained throughout the paper, but for now, only the features of the valves will be discussed. First, notice that the small randomness in valve placement has only a weak effect on the solutions. Regions of closed valves are shaded for clarity. Between two closely spaced closed valves, the flow is small when compared to the flow in regions with many open valves. The pressure has a characteristic step-like pattern in regions of closed valves. The radius displays large discontinuities at the locations of closed valves. The small jumps in $\bar{P}$ and $\bar{R}$ in the open sections are due to the finite open valve resistance. The solid lines closely follow the continuous dashed line which is the valve continuum solution we will present in later sections. For comparison purposes, the solution to the equivalent valveless problem is shown with a dotted line. The goal of the next three subsections will be to understand the cusps and discontinuities in the solid curves and find a way to smooth out the fluid dynamics in regions of open and closed valves to obtain the appropriate valve continuum (dashed curves). First, we will characterize $\bar{P}$, $\bar{Q}$, and $\bar{R}$ in powers of $\epsilon$ in regions where valves are closed; in doing so, we will see that enforcing zero flow and a continuous pressure profile gives a good approximation to our collection of discretely placed closed valves. Then, we show how to obtain a homogenized resistance describing flow through many open valves. Throughout the rest of the paper, constant valve spacing will be assumed $\bar{x}_v^i = i \epsilon$ for simplicity. 

\subsection{Closely spaced closed valves suppress flow} \label{sec:closed_valves}
\begin{figure*}
    \centering
    \includegraphics[width=.85\textwidth]{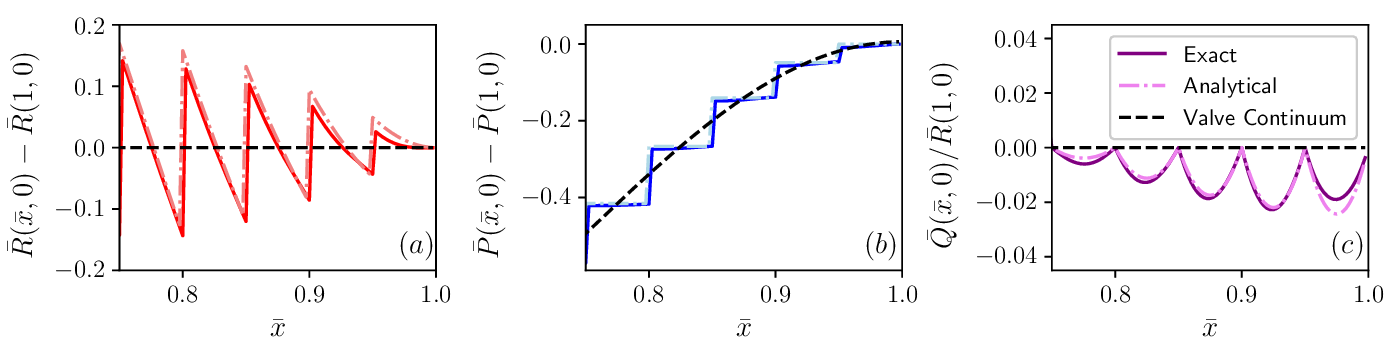}
    \caption{Exact numerical solutions to \eqref{eq:continuity}, \eqref{eq:momentum}, \eqref{eq:PR}, and \eqref{eq:valve} with $f(\bar{x},\bar{t})=\cos(2\pi(\Bar{x}-\bar{t}))$ in a region of many closed valves are shown with solid lines. The analytical predictions \eqref{eq:P_many_valves}, \eqref{eq:Q_many_valves}, \eqref{eq:R_many_valves} are shown with dash-dotted lines, and appear to agree well with the exact solutions. The valve continuum solutions \eqref{eq:P_closed_VC}, \eqref{eq:Q_closed_VC}, and \eqref{eq:R_closed_VC} are shown with black dashed lines. Parameters used for this simulation are $n_v=20$, $\bar{r}_v=0$, $\kappa = .5$, and $\eta_P=.5$. Subtracting (or in the case of the flow, dividing) by a constant value eliminates the unknown parameters $\Bar{R}_0$ and $\Bar{P}_1$. In the case of the analytical solutions, the pressure and radius should be evaluated at the limit as $\bar{x}$ approaches one from below.}
    \label{fig:closed_region}
\end{figure*}

The flow between two closed valves spaced a distance $x_v$ apart is identically the flow in a flexible pipe of length $x_v$ capped at both ends. When $x_v \ll \lambda$, peristalsis is nearly synchronous across the entire pipe, and $\epsilon$ can be used as an expansion parameter. We change to using spatial coordinate $\bar{y}\equiv \bar{x}/\epsilon$ such that the fluid between valves $i$ and $i+1$ is confined to an interval of length one, $\bar{y} \in [\bar{y}_v^i, \bar{y}_v^{i+1}]$. Since $f$ is nearly constant in space between two closed valves, we Taylor expand $f$ about the midpoint between two valves $\Bar{x}_m^i \equiv \frac{1}{2}(\Bar{x}_v^i +\Bar{x}_v^{i+1}) $. 
\begin{align*}
    f(\bar{y}, \bar{t}) &= f (\Bar{x}_m^i,\bar{t}) + \epsilon f'(\Bar{x}_m^i,\bar{t}) (\bar{y}- \bar{y}_m^i)\\ & \hspace{.4cm}+ \epsilon^2 \frac{1}{2}f'' (\Bar{x}_m^i,\bar{t}) (\bar{y}-\bar{y}_m^i)^2+...
\end{align*}
Here, the primes denote derivatives with respect to $\bar{x}$. Equations \eqref{eq:continuity} and \eqref{eq:momentum} become
\begin{equation*}
    \frac{\partial \bar{Q}}{\partial \bar{y}} +\epsilon  \frac{\partial \bar{R}^2}{\partial \bar{t}} = 0 ,
\end{equation*}
\begin{equation*}
    \epsilon \bar{Q} = -\bar{R}^4 \frac{\partial \bar{P}}{\partial \bar{y}},
\end{equation*}
which along with \eqref{eq:PR} can be solved perturbatively with zero flow boundary conditions. We expand $\Bar{P}(\bar{y},\bar{t}) = \Bar{P}_0(\bar{y},\bar{t}) + \epsilon \bar{P}_1(\bar{y},\bar{t})+ \epsilon^2 \bar{P}_2(\bar{y},\bar{t})+...$, and similarly for $\Bar{Q}(\bar{y},\bar{t})$ and $\Bar{R}(\bar{y},\bar{t})$. To lowest order, we have
\begin{equation*}
    \frac{\partial \bar{Q}_0(\Bar{y},\Bar{t})}{\partial \bar{y}} = 0, 
\end{equation*}
\begin{equation*}
    0 = -\bar{R}_0(\Bar{y},\Bar{t})^4 \frac{\partial \bar{P}_0(\Bar{y},\Bar{t})}{\partial \bar{y}},
\end{equation*}
\begin{equation*}
    \bar{P}_0(\Bar{y},\Bar{t})= \kappa (\bar{R}_0(\Bar{y},\Bar{t})-1) + \eta_P f (\bar{x}_m^i,\bar{t})  .
\end{equation*}
From the first equation, the flow is constant in $\bar{y}$, and the boundary conditions fix that constant to zero. From the second equation, the pressure is independent of $\bar{y}$, and the last equation implies that the radius must also be independent of $\bar{y}$. The first-order equations read 
\begin{equation*}
    \frac{\partial \bar{Q}_1(\Bar{y},\Bar{t})}{\partial \bar{y}} +\frac{\partial \bar{R}_0^2(\Bar{t})}{\partial \bar{t}} = 0 ,
\end{equation*}
\begin{equation*}
    0 = -\bar{R}_0(\Bar{t})^4 \frac{\partial \bar{P}_1(\Bar{y},\Bar{t})}{\partial \bar{y}} ,
\end{equation*}
\begin{equation*}
    \bar{P}_1(\Bar{y},\Bar{t}) = \kappa \bar{R}_1 (\Bar{y},\Bar{t}) + \eta_P f'(\bar{x}_m^i,\bar{t}) (\bar{y}- \bar{y}_m^i).
\end{equation*}
The first equation permits flow which is linear in $\Bar{y}$, but our boundary conditions forbid this unless $\Bar{Q}_1 = 0$, which implies $\Bar{R}_0$ is also independent of time. The second equation implies $\Bar{P}_1$ is independent of $\Bar{y}$, and its time dependence is related to $\bar{R}_1$ by the third equation. The second order equations read 
\begin{equation*}
    \frac{\partial \bar{Q}_2(\Bar{y},\Bar{t})}{\partial \bar{y}} + 2\bar{R}_0 \frac{\partial \bar{R}_1(\Bar{y},\Bar{t})}{\partial \bar{t}} = 0, 
\end{equation*}
\begin{equation*}
     0 = -\bar{R}_0^4 \frac{\partial \bar{P}_2(\Bar{y},\Bar{t})}{\partial \bar{y}} ,
\end{equation*}
\begin{equation*}
    \bar{P}_2 (\Bar{y},\Bar{t}) = \kappa \bar{R}_2 (\Bar{y},\Bar{t}) + \frac{1}{2}\eta f'' (\bar{x}_m^i,\bar{t}) (\bar{y}-\bar{y}_m^i)^2.
\end{equation*}
The second equation tells us that $\bar{P}_2$ is independent of $\bar{y}$, and its time dependence is related to $\Bar{R}_2$ by the third equation. Integrating the first equation from $\bar{y}_m-\frac{1}{2}$ to $\bar{y}_m+\frac{1}{2}$ gives the constraint
\begin{equation*}
    \int_{\bar{y}_m-\frac{1}{2}}^{\bar{y}_m+\frac{1}{2}} \frac{\partial \Bar{R}_1(\Bar{y},\bar{t})}{\partial \Bar{t}} d\Bar{y} = \frac{1}{\kappa}\frac{\partial \Bar{P}_1(\Bar{t})}{\partial \Bar{t}} = 0.
\end{equation*}
So, in fact, $\Bar{P}_1$ is a constant. The continuity equation and boundary conditions are satisfied by
\begin{equation*}
    \Bar{Q}_2(\Bar{y}, \Bar{t}) - \bar{R}_0 \frac{\eta_P}{\kappa}   \frac{\partial f'(\Bar{x}_m, \Bar{t})}{\partial \Bar{t}}(\Bar{y}-\Bar{y}_v^i)(\Bar{y}-\Bar{y}_v^{i+1})= 0.
\end{equation*}
Assuming the valves are equally spaced, we can make the replacement $\bar{x}_m^i = \epsilon \lfloor \Bar{x}/\epsilon \rfloor+\epsilon/2$, and write an expression for the pressure, flow, and radius:
\begin{widetext}
\begin{align}
    \bar{P}(\bar{x}, \bar{t}) = \kappa(\bar{R}_0 - 1) +\eta_P f\left(\epsilon \lfloor \Bar{x}/\epsilon  \rfloor+\epsilon/2,\bar{t}\right) + \epsilon \bar{P}_1 + \epsilon^2 \Bar{P}_2(\bar{t}) + O(\epsilon^3) ,\label{eq:P_many_valves}
\end{align}
\begin{align}
    \bar{Q}(\bar{x}, \bar{t})= \frac{\eta_P }{\kappa}\bar{R}_0 \frac{\partial f\left(\epsilon \lfloor \Bar{x}/\epsilon  \rfloor+\epsilon/2,\bar{t}\right)}{\partial \bar{t}} \left(\bar{x}- \epsilon \Big{\lfloor} \frac{\Bar{x}}{\epsilon} \Big{\rfloor}\right)\left(\bar{x}- \epsilon \Big{\lfloor} \frac{\Bar{x}}{\epsilon} \Big{\rfloor}-\epsilon\right)+O(\epsilon^3) ,\label{eq:Q_many_valves}
\end{align}
\begin{align}
    \bar{R}(\bar{x}, \bar{t}) = \bar{R}_0 &+  \frac{\epsilon}{\kappa}\Bar{P}_1 - \frac{\eta_P}{\kappa} f'\left(\epsilon \lfloor \Bar{x}/\epsilon  \rfloor+\epsilon/2,\bar{t}\right)\left(  \bar{x}- \epsilon \Big{\lfloor} \frac{\Bar{x}}{\epsilon} \Big{\rfloor}-\frac{\epsilon}{2} \right) \nonumber \\&+ \frac{\epsilon^2}{\kappa}\Bar{P}_2(\bar{t}) - \frac{1}{2}\frac{\eta_P}{\kappa} f''\left(\epsilon \lfloor \Bar{x}/\epsilon  \rfloor+\epsilon/2,\bar{t}\right) \left(  \bar{x}- \epsilon \Big{\lfloor} \frac{\Bar{x}}{\epsilon} \Big{\rfloor}-\frac{\epsilon}{2} \right)^2 +O(\epsilon^3). \label{eq:R_many_valves}
\end{align}
\end{widetext}
Here, $\bar{R}_0$ and $\Bar{P}_1$ are undetermined constants and $\Bar{P}_2(\Bar{t})$ is an undetermined function of time. These approximate expressions for a confined fluid agree well with the exact solutions, as demonstrated in figure \ref{fig:closed_region}. 
The pressure is nearly constant in space in between two valves, and the time dependence is dominated by the value of $f(\bar{x}_v^i,\bar{t})$, explaining the step-like profiles observed in figure \ref{fig:closed_region}$b$. The flow is everywhere continuous, but cusps can be seen at valves, as shown in figure \ref{fig:closed_region}$c$. Crucially, the flow between the valves is suppressed by $\epsilon^2$, so the induced flow decreases rapidly as the valve spacing decreases, but may be noticeable in a highly compliant vessel with small $\kappa$. The radius displays rapid oscillations due to the $\bar{y}$-dependent term at $O(\epsilon^1)$, as seen in figure \ref{fig:closed_region}$a$. These oscillations describe how slightly different forces applied between two valves can cause a large gradient in the radius.

Continuous equations can be recovered by taking the limit $\epsilon \rightarrow 0$:
\begin{equation}
    \lim_{\epsilon \rightarrow 0}\bar{P}(\bar{x}, \bar{t}) = \kappa(\bar{R}_0 - 1) +\eta_P f(\bar{x},\bar{t}) ,\label{eq:P_closed_VC}
\end{equation}
\begin{equation}
     \lim_{\epsilon \rightarrow 0} \bar{Q}(\bar{x}, \bar{t}) =0 ,\label{eq:Q_closed_VC}
\end{equation}
\begin{equation}
     \lim_{\epsilon \rightarrow 0} \bar{R}(\bar{x}, \bar{t}) = \bar{R}_0. \label{eq:R_closed_VC}
\end{equation}
These equations describe the valve continuum in a region of closed valves and can be understood as a simple consequence of the zero flow boundary conditions. Not only does this enforce $\bar{Q}=0$, but also the radius must be kept constant to prevent induced flow. With this idea in mind, note that we cannot apply an arbitrary radius-imposed peristalsis in between two closed valves because doing so would violate the zero flow condition. The model \citep{Farina_Fusi_Fasano_Ceretani_Rosso_2016} utilizing ideal valves and radially imposed peristalsis considers the case of only two valves, and they find that at least one of the valves must be open at any time. We will later see that it is possible to have radius-imposed peristalsis through many valves, but only if precisely one valve is closed per wavelength.

\subsection{Flow through many open valves} \label{sec:open_valves}
Next, we seek a simplified model for a region with many open valves. Notice that we can incorporate open valves into our one-dimensional momentum equation by introducing an additional resistance $\bar{r}_v$ at the location of each valve:
\begin{equation}
    \frac{\partial \bar{P}}{\partial \bar{x}} = -\bar{R}^{-4} \Big[1+ \bar{r}_v \sum_i \delta(\bar{x}-\bar{x}_v^i)  \Big] \bar{Q}. \label{eq:momentum_modified}
\end{equation}
The calculation will be similar to the previous section, but here there will be two length scales which are important. The valves change the fluidic resistance over a small length scale $\bar{y}\equiv \Bar{x}/\epsilon$ while the channel changes the resistance over a longer length scale $\Bar{x}$, so we can apply the tools of homogenization theory to derive an appropriate effective resistance \citep{Holmes_2013}. We will expand each of the functions $\bar{P}, \bar{Q}, \bar{R}$ in powers of $\epsilon$, and introduce $n_v = \epsilon^{-1}$ which counts the number of valves per wavelength, with $n_v \bar{r}_v \sim O(\epsilon^0)$. In a region of open valves, 
\begin{widetext}
\begin{equation}
   \epsilon \Big( \frac{\partial}{\partial \bar{x}} + \frac{1}{\epsilon} \frac{\partial}{\partial \bar{y}} \Big) \Big(\bar{Q}_0 + \epsilon \bar{Q}_1 + ...\Big) + \epsilon \frac{\partial}{\partial \bar{t}} \Big(\bar{R}_0^2 + ...\Big) = 0, \label{eq:continuity_modified2}
\end{equation}
\begin{equation}
    \epsilon \Big(\frac{\partial}{\partial \bar{x}} + \frac{1}{\epsilon} \frac{\partial}{\partial \bar{y}} \Big)\Big(\bar{P}_0 + \epsilon \bar{P}_1 + ...\Big)= - \epsilon \Big(\bar{R}_0^{-4}+...\Big) \Big[1+n_v \bar{r}_v \sum_i \delta(\bar{y}-\bar{y}_v^i) \Big] \Big(\bar{Q}_0 + ... \Big), \label{eq:momentum_modified2}
\end{equation}
\begin{equation}
    \Big(\bar{P}_0 + \epsilon \bar{P}_1 + ...\Big) = \kappa \left[\Big(\bar{R}_0 + \epsilon \bar{R}_1 + ...\Big) - 1 \right]+ \eta_P f .\label{eq:PR_modified2}
\end{equation}
\end{widetext}
From the $O(\epsilon^{0})$ terms in \eqref{eq:continuity_modified2} and \eqref{eq:momentum_modified2}, we immediately learn that $\bar{Q}_0$ and $\bar{P}_0$ are independent of $\bar{y}$. From \eqref{eq:PR_modified2}, we can also see that $\bar{R}_0$ is independent of $\bar{y}$. 
At $O(\epsilon^1)$ in equation \eqref{eq:PR_modified2}, we have
\begin{equation}
    \bar{P}_0 = \kappa(\bar{R}_0 - 1) + \eta_P f. \label{eq:PR_modified2_0}
\end{equation}
At $O(\epsilon^1)$ in equation \eqref{eq:continuity_modified2}, we have
\begin{equation*}
     \frac{\partial \bar{Q}_0(\Bar{x},\Bar{t})}{\partial \bar{x}} +  \frac{\partial \bar{Q}_1(\Bar{x},\Bar{y},\Bar{t})}{\partial \bar{y}}  + \frac{\partial \bar{R}_0(\Bar{x},\Bar{t})^2}{\partial \bar{t}} = 0.
\end{equation*}
We see that $\Bar{Q}_1$ is linear in $\Bar{y}$, but since $\Bar{y}$ describes effects localized to the valves, we must have that $\bar{Q}_1$ does not grow far from the valves, and thus $\frac{\partial \bar{Q}_1(\Bar{x},\Bar{y},\Bar{t})}{\partial \bar{y}} = 0$. This leaves us with a continuity equation purely in terms of $\bar{x}$:
\begin{equation}
     \frac{\partial \bar{Q}_0}{\partial \bar{x}} +\frac{\partial \bar{R}_0^2}{\partial \bar{t}} = 0 .\label{eq:continuity_modified2_0}
\end{equation}
Looking at $O(\epsilon^1)$ in equation \eqref{eq:momentum_modified2}, we have
\begin{widetext}
\begin{align*}
    \frac{\partial \bar{P}_0(\Bar{x},\Bar{t})}{\partial \bar{x}} + \frac{\partial \bar{P}_1(\Bar{x},\Bar{y},\Bar{t})}{\partial \bar{y}} = -\bar{R}_0(\Bar{x},\Bar{t})^{-4} \Big[1+n_v \bar{r}_v \sum_i \delta(\bar{y}-\bar{y}_v^i) \Big] \bar{Q}_0(\Bar{x},\Bar{t}).
\end{align*}
Integrating over $\Bar{y}$ from an arbitrary point $\bar{y}_0$ to $\Bar{y}$ gives, 
\begin{align*}
    \Bar{P}_1(\Bar{x},\Bar{y},\Bar{t}) - \Bar{P}_1(\Bar{x},\Bar{y}_0,\Bar{t}) &=  - \int_{\bar{y}_0}^{\bar{y}}\left[\frac{\partial \bar{P}_0}{\partial \bar{x}}+\bar{R}_0^{-4} \Big[1+ n_v \bar{r}_v \sum_i \delta(\bar{y}'-\bar{y}_v^i) \Big] \bar{Q}_0\right]d\Bar{y}'\\
    &=- \left( \frac{\partial \bar{P}_0}{\partial \bar{x}}+\bar{R}_0^{-4} \bar{Q}_0\right) (\bar{y}-\bar{y}_0) - n_v \bar{r}_v \bar{R}_0^{-4} \bar{Q}_0 \int_{\Bar{y}_0}^{\bar{y}}\sum_i \delta(\bar{y}'-\bar{y}_v^i)d\bar{y}'. 
\end{align*}
Each of the terms on the right-hand side diverges for large $\Bar{y}$, so the only way for $\Bar{P}_1$ to remain finite is to have these terms cancel as $\Bar{y}$ increases
\begin{equation*}
    \lim_{\bar{y}\rightarrow \infty} \frac{1}{\Bar{y}-\bar{y}_0} \left[  - \left( \frac{\partial \bar{P}_0}{\partial \bar{x}}+\bar{R}_0^{-4} \bar{Q}_0\right) (\bar{y}-\bar{y}_0) - n_v \bar{r}_v \bar{R}_0^{-4} \bar{Q}_0 \int_{\Bar{y}_0}^{\bar{y}}  \sum_i \delta(\bar{y}'-\bar{y}_v^i)d\bar{y}\right] = 0.
\end{equation*}
\end{widetext}
The integral in the second term grows like $\bar{y}-\bar{y}_0$ for large $\Bar{y}$, so our homogenized momentum equation simply becomes 
\begin{equation}
    \frac{\partial \bar{P}_0}{\partial \bar{x}} = -\bar{R}_0^{-4} \left[ 1+n_v \Bar{r}_v \right]\bar{Q}_0 \label{eq:homogenized}.
\end{equation}
This equation could have been guessed by adding the resistance of our tube and $n_v$ valves in series. In the limit $\epsilon \rightarrow 0 $, equations \eqref{eq:PR_modified2_0}, \eqref{eq:continuity_modified2_0}, and \eqref{eq:homogenized} are exact. These are the expressions for the valve continuum in a region of open valves. While $\bar{P}_1$ and $\bar{R}_1$ are both discontinuous at the valves, $\bar{Q}_1$ is independent of the microscopic coordinate $\bar{y}$, so much like the theory for regions of closed valves, the leading-order term in our expansion for regions of open valves works particularly well at describing the flow which is the fundamental quantity of interest. 

The dependence on valve parameters can be eliminated entirely by rescaling the amplitude and stiffness according to $\eta_P \rightarrow \eta_P/(1+n_v \Bar{r}_v)$ and $\kappa \rightarrow \kappa/(1+n_v \Bar{r}_v)$. This will give the correct radius and flow, but the result for the pressure will then need to be multiplied by a factor of $(1+n_v \Bar{r}_v)$ to get the correct homogenized pressure. This procedure was done to obtain the valve continuum solutions (solid lines) in figure \ref{fig:convergence}. For notational simplicity, we will simply set $\Bar{r}_v=0$ until we are interested in exploring specific features related to the number of valves. 

Throughout the next section, all analytical results are presented in the limit $\epsilon \rightarrow 0$, and the subscript zero will be re-purposed for a new perturbative expansion. 

\subsection{Matching conditions}\label{sec:matching}
So far, we have derived the equations for a continuum of open valves and a continuum of closed valves. What remains is to match these regions. Matching occurs at coordinates where the valves are closing or opening. From equations \eqref{eq:closes} and \eqref{eq:opens}, we know that the pressure must be continuous at these coordinates, and \eqref{eq:force_balance} implies the radius is also continuous. Just as in the discrete valve system, the flow is always continuous and is identically zero during opening or closing. By \eqref{eq:homogenized}, the pressure gradient in an open region adjacent to a closed region must be zero to ensure zero flow at the transition, but the pressure gradient in a closed region adjacent to an open region only needs to be nonnegative, so the pressure gradient need not be continuous. Thus, although $P, Q$, and $R$ are continuous in the valve continuum, the functions may not be smooth at the closing and opening coordinates. This behavior can be seen in the valve continuum predictions in figure \ref{fig:convergence} where cusps can be seen where the shaded and unshaded regions match. Later, we will show that for the case of traveling waves, the pressure gradient is continuous at the closing coordinates but not at the opening coordinates.

To summarize, a fluid in an elastic pipe containing many valves with arbitrary imposed force $f$ satisfies equations \eqref{eq:P_closed_VC}, \eqref{eq:Q_closed_VC}, and \eqref{eq:R_closed_VC} in a region containing many closed valves, and \eqref{eq:PR_modified2_0}, \eqref{eq:continuity_modified2_0}, and \eqref{eq:homogenized} in a region containing many open valves. These regions are matched by continuity of $P,Q,$ and $R$. This is the general form of the valve continuum. Additional boundary conditions could be imposed on a finite tube. We will only consider $f$ in the form of forward-propagating and backward-propagating peristaltic waves imposed on an infinitely long tube with zero net pressure drop such that periodic boundary conditions in $P, Q,$ and $R$ apply. 

\section{Valve Continuum Results I: Forward-Propagating Peristaltic Waves} \label{sec:valve_continuumFW}
Combining the results of section \ref{sec:valves} gives us the behavior of $\bar{P}$, $\bar{Q}$, and $\bar{R}$ when $\epsilon \rightarrow 0$. The motivation for studying this regime of dense valves is twofold. First, we have eliminated any dependence on the valve positions $\{ \bar{x}_v^i \}$ and by appropriate rescaling have even eliminated $\bar{r}_v$, leaving only two parameters characterizing the peristaltic pumping. Second, because the valves no longer break translation symmetry, we can study peristaltic waves using ODEs in terms of a single wave coordinate. In the next two sections, we will consider forward- and backward-propagating waves, respectively (where valves always promote flow in the positive direction). For the valveless system, these two systems are related trivially by time-reversal symmetry, but the presence of valves will require us to study these solutions separately. It will be convenient when discussing the role of valves to consider only functions $f$ which have a unique local maximum and minimum. This will ensure that within each wavelength, there is one continuous region of open valves (where $\frac{\partial \Bar{P}}{\partial \bar{x}}<0$) and one continuous region of closed valves (where $\frac{\partial \Bar{P}}{\partial \bar{x}}>0$). 

\begin{figure*}
    \centering
    \includegraphics[width=.75\textwidth]{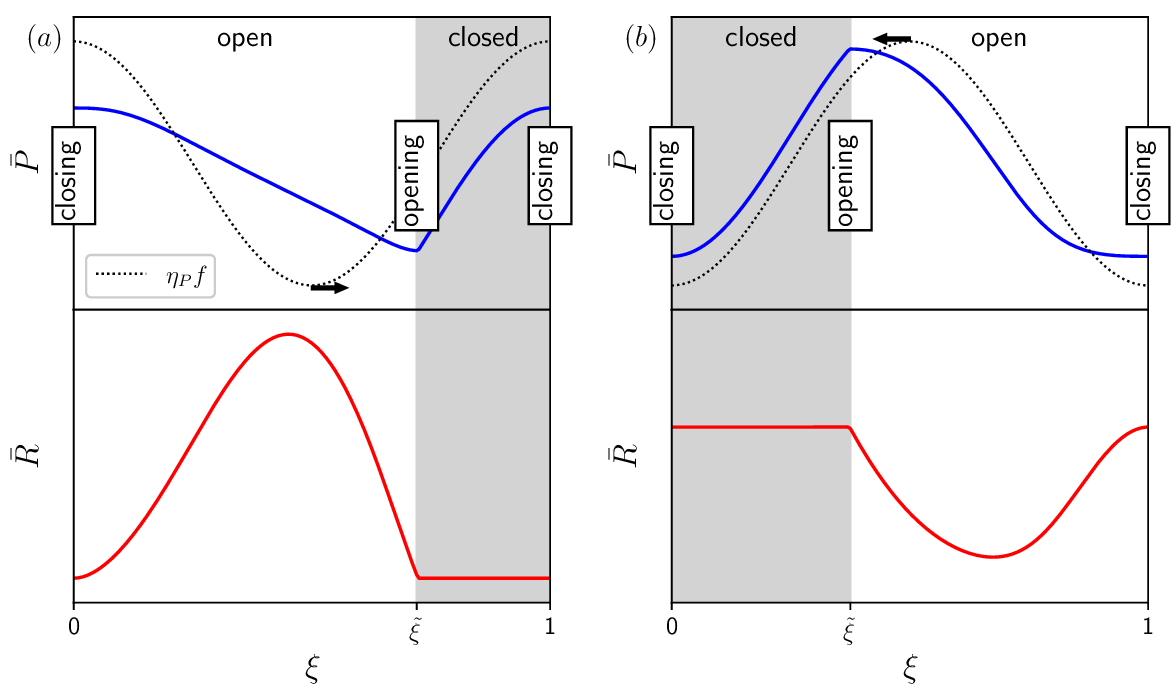}
    \caption{Example solutions to the valve continuum model which demonstrate the appropriate matching conditions. $(a)$ For a forward-propagating wave, closing occurs at $\max \bar{P}$ and $\min \Bar{R}$, so it must also occur at $\max f$ by \eqref{eq:PR}. The origin is chosen to be $\max f$ for simplicity. However, opening occurs at $\min \Bar{P}$ and $\min \bar{R}$, so the opening coordinate  $\Tilde{\xi}$ cannot be simply expressed in terms of $f$. $(b)$ By a similar argument, for a backward-propagating wave, closing occurs at $\min f$, but the opening coordinate $\tilde{\xi}$ cannot be simply expressed in terms of $f$. Note that opening is defined as the time at which the valve transitions from closed to open, and closing is defined as the time at which the valve transitions from open to closed. For a forward-propagating wave where $\xi=\bar{x}-\bar{t}$, one should read the plots from right to left when determining the opening and closing coordinates, but for a backward-propagating wave where $\xi=\bar{x}+\bar{t}$, one should read the plots from left to right when determining the opening and closing coordinates.}
    \label{fig:demo}
\end{figure*}

In this section, we will consider forward-propagating peristaltic forces of the form 
\begin{equation}
    f(\Bar{x},\Bar{t}) = f(\Bar{x}-\Bar{t}) \equiv f(\xi).
\end{equation}
Since all functions now only depend on $\xi$, we will sometimes use primes to unambigiously denote derivatives with respect to $\xi$. The origin is chosen such that closing occurs at $\xi = 0$. We will show that this corresponds to a simple phase shift of $f$. First, let's consider the continuity equation for this model. By our choice of origin, $\Bar{Q}(0)=0$, the continuity equation \eqref{eq:continuity} is simply 
\begin{equation}
    \frac{d }{d \xi} \Big[ \bar{Q} - \bar{R}^2 \Big] = 0 \implies \Bar{Q}(\xi) = \Bar{R}^2(\xi) - \Bar{R}^2(0) .\label{eq:continuityV} 
\end{equation}
From our work in section \ref{sec:closed_valves}, we not only know that the flow in the closed regions will be zero, but also the radius in the closed regions will be constant and equal to $\Bar{R}(0)$. Equation \eqref{eq:continuityV}, along with the fact that $\Bar{Q}(\xi) \geq 0$ enforces that $\Bar{R}(0) = \min \Bar{R}$. Also, the pressure in the open regions is decreasing while the pressure in the closed regions is increasing, so the opening and closing coordinates must be relative extrema. Considering the sign of the pressure gradient at a previous time step reveals that closing occurs at the maximum value of $\Bar{P}$, and opening occurs at the minimum value of $\Bar{P}$. Therefore, at closing, $\eta_P f(0)=\max \Bar{P}-\kappa(\min \Bar{R}-1) = \eta_P \max f$. Thus, we can ensure that the valve closes at the origin by shifting the origin to align with the maximum value of $f$:
\begin{equation}
    f(0) = \max f \iff \text{valve closes at $\xi=0$}.
\end{equation}
A summary of this argument is given in figure \ref{fig:demo}$a$. The fact that we have this simple matching condition during closing has further implications. Since $f(0)=\max f$, assuming $f$ has continuous first derivative, then $f'(0) = 0$. As mentioned in section \ref{sec:matching}, we only require continuity between the closed and open regions, but since $f'(0)=0$, and $\bar{P}'(0^+)=0$ to ensure continuity in the flow, $\bar{R}'(0^+)=0$ by \eqref{eq:PR}, which also implies $\bar{Q}'(0^+)=0$ by \eqref{eq:continuityV}. Similarly, since $f'(1)=0$ and $\bar{R}'(1^-)=0$ in a closed region, \eqref{eq:PR} implies $\bar{P}'(1^-)=0$. Therefore, during closing, the pressure, flow, and radius are not only continuous, but also have a continuous first derivative. 

The opening coordinate, which we denote $\tilde{\xi}$, does not possess a simple form. Attempting to apply a similar argument gives $\eta_P f(\Tilde{\xi}) = \min \Bar{P}-\kappa(\min \Bar{R}-1)$, which cannot be simplified purely in terms of $f$. Since $f'(\Tilde{\xi})$ need not be zero, we typically observe a cusp at $\Tilde{\xi}$. 

It is often useful to combine equations \eqref{eq:Q_closed_VC} and \eqref{eq:homogenized} (with $\Bar{r}_v=0)$ into a single momentum equation that will govern the entire valve continuum: 
\begin{equation}
    \bar{Q} = -\bar{R}^4 \frac{d \bar{P}}{d \xi } \Theta \Big( -\frac{d \bar{P}}{d \xi } \Big). \label{eq:momentumV}
\end{equation}
For regions with open valves, we can decouple \eqref{eq:continuityV}, \eqref{eq:momentumV}, and \eqref{eq:PR} into a single nonlinear ODE. To summarize, the radius is given by solving
\begin{equation}
    \begin{cases}
        \frac{d\bar{R}}{d\xi} =  -\frac{1}{\kappa}  \bar{R}^{-4} \Big(\bar{R}^{2} - \bar{R}(0)^2  \Big) - \frac{\eta_P}{\kappa} \frac{df}{d\xi}  & 0\leq \xi \leq \Tilde{\xi} \\
        \Bar{R}(\xi) = \Bar{R}(0) & \Tilde{\xi} \leq \xi \leq 1 \label{eq:Rvc}
    \end{cases}
\end{equation}
where $\Tilde{\xi} \in (0,1]$ is found by applying continuity of $\Bar{R}$ during opening, and the constant $\Bar{R}(0)$ is fixed by enforcing volume conservation \eqref{eq:RSquaredConstraint}, which for the valve continuum model with forward-propagating peristaltic waves takes the special form
\begin{equation}
    \int_0^{\Tilde{\xi} } \Bar{R}^2(\xi) d\xi + (1-\tilde{\xi}) \Bar{R}^2(0) = 1. \label{eq:RSquaredConstraintVC}
\end{equation}
Taking the average of \eqref{eq:continuityV} and applying \eqref{eq:RSquaredConstraintVC} gives a simple expression for the mean flow:
\begin{align}
    \langle \bar{Q} \rangle
 &= 1 - \Bar{R}^2(0). \label{eq:QbarVC_FW}
\end{align}
The differential equation for the open region looks identical to that for the valveless problem. However, the boundary conditions make our problem significantly harder to solve. During the remainder of the section, specific regimes will be studied analytically and numerically. 

\subsection{Forward-Propagating Peristalsis in a Stiff Tube (Radius-imposed peristalsis)} \label{sec:stiffFW}
\begin{figure*}
    \centering
    \includegraphics[width=.75\textwidth]{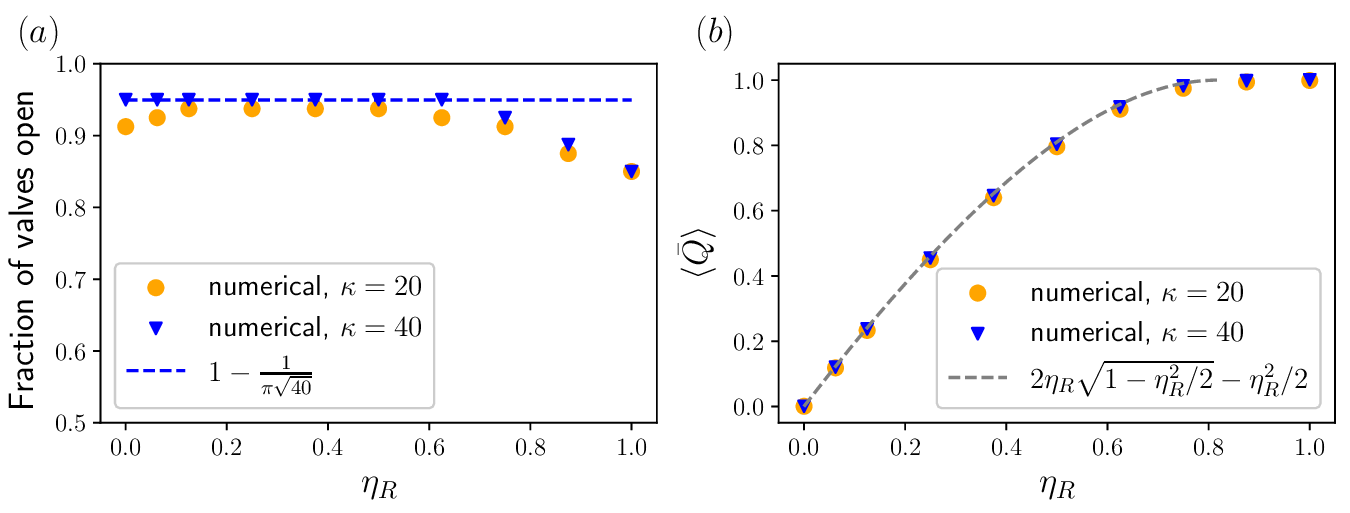}
    \caption{Results for forward-propagating peristalsis in a stiff tube with a continuum of valves, $f(\bar{x},\Bar{t}) = \cos(2\pi (\Bar{x}-\Bar{t}))$. $(a)$ Fraction of valves open in a stiff vessel for two different choices of large stiffness and varying radial amplitude $\eta_R$. The dashed line is the small-amplitude analytic result \eqref{eq:xiTilde1_Stiff}. $(b)$ Mean flow for two different choices of large stiffness and varying radial amplitude $\eta_R$. The dashed line is the analytic result \eqref{eq:QbarStiffFW}. }
    \label{fig:stiffVC}
\end{figure*}

Taking the limit of an infinitely stiff vessel $\kappa \rightarrow \infty$ in equation \eqref{eq:Rvc} will lead to the trivial result of zero radial deformation and zero induced flow. However, if we also consider a large amplitude of (force-imposed) peristalsis $\eta_P$ such that $\eta_R \equiv \eta_P/\kappa$ is finite but small, then we recover radius-imposed peristalsis with 
\begin{equation}
    \Bar{R}(\xi) = \sqrt{1-\eta_R^2 \langle f^2 \rangle} - \eta_R f(\xi) \label{eq:RstiffFW}
\end{equation}
where $\eta_R$ is the characteristic radial deformation. Initially, this appears problematic because the radius in the closed regions needs to be constant, but \eqref{eq:RstiffFW} is everywhere proportional to $f$. Furthermore, our continuity equation in the open region is a first-order differential equation with two zero flow boundary conditions at the locations which connect to the closed regions. The resolution is to have only a single closed valve per wavelength at $\xi=1$. Indeed, the valve opens at the coordinate $\tilde{\xi}$ satisfying $\Bar{R}(\tilde{\xi}) = \Bar{R}(0)$ which in this case implies $f(\tilde{\xi})=f(0)$, and since we are assuming a unique maximum in $f$, it must be that $\Tilde{\xi}=1$. The pressure will be discontinuous across the closed valve, but the flow is given by 
\begin{align}
    \bar{Q}(\xi) &=  -2 \eta_R \sqrt{1-\eta_R^2 \langle f^2 \rangle} \Big(f(\xi) -\max (f) \Big)\nonumber \\&\hspace{.4cm} +\eta_R^2 \left(f(\xi)^2 - (\max (f) )^2\right),
\end{align}
and its time average is 
\begin{align}
    \langle \Bar{Q} \rangle &= 2 \eta_R \max (f)  \sqrt{1-\eta_R^2 \langle f^2 \rangle} \nonumber \\&\hspace{.4cm} -\eta_R^2 \left((\max (f) )^2 - \langle f^2 \rangle \right) .\label{eq:QbarStiffFW}
\end{align}
This solution is valid only if the radius remains positive which is true provided $\eta_R$ is sufficiently small: 
\begin{equation}
    \eta_R < \frac{1}{\sqrt{(\max f)^2 + \langle f^2 \rangle} }. \label{eq:etaRconstraint}
\end{equation}
For larger $\eta_R$, the tube is completely occluded with all of the fluid volume transported in one period, which in dimensionless units is $\langle \bar{Q} \rangle =1$. Figure \ref{fig:stiffVC} compares the numerical results of the fraction of valves open and the mean flow to the analytic predictions in the case of sinusoidal peristalsis. The fraction of valves open approaches a number close to one, which will be quanitified in the next section. The flow for $\eta_R < \sqrt{2/3}$ is correctly predicted by \eqref{eq:QbarStiffFW}, and is equal to one for larger values of $\eta_R$. 

The flow in this limit is drastically different from that of radius-imposed peristalsis without valves. Perhaps the most striking feature is that the leading-order flow for small-amplitude peristalsis scales with $\eta_R$ in the valve continuum, as opposed to $\eta_R^2$ for the case without valves. It is also worth appreciating that the exact flow possesses an even simpler form than the case of radius-imposed peristalsis without valves which can only be written in terms of integrals of powers of $\Bar{R}$:
\begin{equation}
    \langle \Bar{Q}^{\text{nv}} \rangle = 1 - \frac{\langle \Bar{R}^{-2}\rangle}{\langle \Bar{R}^{-4}\rangle}.
\end{equation}
The notation ``nv'' will be used to denote ``no valves". This equation is easily obtained by integrating $\Bar{Q}\Bar{R}^{-4}$ over one period and applying continuity of pressure. But in the case with ideal valves, the pressure across a single closed valve is discontinuous, and the form is instead fixed by the constraint of zero flow at the valve. Interestingly, the valve continuum solution in this case did not even rely on the precise form of the momentum equation, only that there is precisely one closed valve. 

Sample solutions in this regime are given in the left columns of figures \ref{fig:convergence}$a$ and $b$. For a finite $\kappa$, the fraction of valves open is less than one, and a continuous pressure is observed in the closed region which interpolates between that in the open regions. The pressure for the valveless case is small and nearly averages to zero, but a large negative pressure gradient is established for the case with valves. The flow curve has a similar shape as that of the valveless solution but is shifted upward. 

\subsection{Forward-Propagating, Small-Amplitude Peristalsis}\label{sec:Small_Amplitude_FW}
The opposite limit of vanishing stiffness ends up being a hard problem to solve analytically, but we have already discovered interesting solutions at order $\eta_P$, so we will now investigate the leading-order small-amplitude response for arbitrary $\kappa$.  We attempt to solve equation \eqref{eq:Rvc} up to order $\eta_P$ by expanding $\Bar{R}(\xi) = 1 + \eta_P \Bar{R}_1(\xi)+...$
\begin{equation}
    \begin{cases}
        \frac{d\bar{R}_1}{d\xi} =  -\frac{2}{\kappa} \left(\bar{R}_1-\bar{R}_1(0)\right) - \frac{1}{\kappa}\frac{df}{d\xi}  & 0\leq \xi \leq \Tilde{\xi}_1 \\
        \Bar{R}_1(\xi) = \Bar{R}_1(0) & \Tilde{\xi}_1 \leq \xi \leq 1.
    \end{cases} \label{eq:R1vc}
\end{equation}
Note that $\Tilde{\xi}_1$ is defined as the coordinate at which the open and closed radial solutions at order $\eta_P$ match. The constraint \eqref{eq:RSquaredConstraintVC} becomes
\begin{align*}
    0 &= \int_0^{\tilde{\xi}_1} \Bar{R}_1(\xi) d\xi + (1-\Tilde{\xi}_1) \Bar{R}_1(0)\\
    &= \int_0^{\tilde{\xi}_1} \left[-\frac{\kappa}{2} \frac{d\Bar{R}_1}{d\xi} +\Bar{R}_1(0)-\frac{1}{2}\frac{df}{d\xi} \right] d\xi + (1-\Tilde{\xi}_1) \Bar{R}_1(0)\\
    &= \Bar{R}_1(0) - \frac{1}{2}[f(\tilde{\xi}_1)-f(0)]
\end{align*}
\begin{equation}
    \implies \Bar{R}_1(0) = -\frac{1}{2}[f(0)-f(\tilde{\xi}_1)]
\end{equation}
The solution to \eqref{eq:Rvc} with this initial condition is
\begin{equation}
     \bar{R}_1(\xi) =  -\frac{1}{2}\left[ f(0) - f(\tilde{\xi}_1)  \right] - \frac{1}{\kappa} \int_0^\xi d\xi' \frac{df(\xi')}{d\xi} e^{\frac{2}{\kappa} (\xi'-\xi)}
\end{equation}
where $\tilde{\xi}_1$ satisfies
\begin{equation}
    \int_0^{\Tilde{\xi}_1} d\xi' \frac{df(\xi')}{d\xi} e^{\frac{2}{\kappa} (\xi'-\Tilde{\xi}_1)} = 0. \label{eq:xiTilde}
\end{equation}
We now have $\bar{R}_1$ and can easily write down $\bar{P}_1$ and $\bar{Q}_1$:
\begin{widetext}
\begin{equation}
    \bar{R}_1(\xi)=
    \begin{cases}
      -\frac{1}{2} [f(0)-f(\Tilde{\xi}_1)] - \frac{1}{\kappa} \int_0^\xi d\xi' \frac{df}{d\xi} e^{\frac{2}{\kappa} (\xi'-\xi)}& \text{for } 0\leq \xi \leq \Tilde{\xi}_1 \\
      -\frac{1}{2} [f(0)-f(\Tilde{\xi}_1)]& \text{for } \Tilde{\xi}_1\leq \xi \leq 1 ,
    \end{cases} \label{eq:Rxi}
\end{equation}
\begin{equation}
    \bar{P}_1(\xi)=
    \begin{cases}
        -\frac{ \kappa}{2} [f(0)-f(\Tilde{\xi}_1)] + f(\xi) - \int_0^\xi d\xi' \frac{df}{d\xi} e^{\frac{2}{\kappa} (\xi'-\xi)}& \text{for } 0\leq \xi \leq \Tilde{\xi}_1 \\
        -\frac{\kappa}{2} [f(0)-f(\Tilde{\xi}_1)] + f(\xi) & \text{for } \Tilde{\xi}_1\leq \xi \leq 1 ,
    \end{cases} \label{eq:Pxi}
\end{equation}
\begin{equation}
    \bar{Q}_1(\xi)=
    \begin{cases}
      - \frac{2 }{\kappa}  \int_0^\xi d\xi' \frac{df}{d\xi} e^{\frac{2}{\kappa}(\xi'-\xi)}& \text{for } 0\leq \xi \leq \Tilde{\xi}_1 \\
      0& \text{for } \Tilde{\xi}_1\leq \xi \leq 1 .
    \end{cases} \label{eq:Qxi}
\end{equation}
\end{widetext}
Clearly $\kappa$ plays an important role in the shape of these curves, but this will be easier to understand once an explicit solution is given in the next section. 

Additionally, we can calculate the mean flow in terms of $\Tilde{\xi}_1$:
\begin{align}
    \langle \bar{Q}_1 \rangle &= -2 \bar{R}_1(0) = f(0)-f(\Tilde{\xi}_1) .\label{eq:Qbar}
\end{align}
Evaluating this simple-looking equation requires knowing $\Tilde{\xi}_1$ which depends on $\kappa$. We will now show these results for the special case of a sine wave $f(\xi) = \cos(2\pi \xi)$: 
\begin{equation}
    e^{-\frac{2}{\kappa} \Tilde{\xi}_1} - \cos (2\pi \Tilde{\xi}_1) + \frac{1}{\pi \kappa} \sin (2\pi \Tilde{\xi}_1) = 0 .\label{eq: transcendental}
\end{equation}
\begin{widetext}
\begin{equation}
   \bar{R}_1(\xi)=
    \begin{cases}
       -\frac{1}{2} \left[1-\cos(2\pi \tilde{\xi}_1)\right]  + \pi \left[ \frac{\pi \kappa e^{-\frac{2}{\kappa}\xi} + \sin (2\pi \xi)- \pi \kappa \cos (2\pi \xi) }{1+(\pi \kappa)^{2}}\right] ,& \text{for } 0\leq \xi \leq \Tilde{\xi}_1 \\
       -\frac{1}{2} \left[1-\cos(2\pi \tilde{\xi}_1)\right] ,& \text{for } \Tilde{\xi}_1\leq \xi \leq 1 
    \end{cases}\label{eq:Rsine}
\end{equation}
\begin{equation}
    \bar{P}_1(\xi)=
    \begin{cases}
        -\frac{\kappa}{2} \left(1-\cos(2\pi \tilde{\xi}_1)\right)+ \frac{(\pi \kappa)^2 e^{-\frac{2}{\kappa}\xi} + \cos (2\pi \xi) + \pi \kappa \sin (2\pi \xi)}{1+(\pi \kappa)^{2}} ,& \text{for } 0\leq \xi \leq \Tilde{\xi}_1 \\
        -\frac{\kappa}{2} \left(1-\cos(2\pi \tilde{\xi}_1)\right)+ \cos (2\pi \xi),& \text{for } \Tilde{\xi}_1 \leq \xi \leq 1 
    \end{cases}\label{eq:Psine}
\end{equation}
\begin{equation}
   \bar{Q}_1(\xi) =
    \begin{cases}
      2\pi  \left[\frac{\pi \kappa e^{-\frac{2}{\kappa} \xi} +  \sin (2\pi \xi) - \pi \kappa \cos (2\pi \xi) }{1+(\pi \kappa)^{2}}\right],& \text{for } 0\leq \xi \leq \Tilde{\xi} \\
      0,& \text{for } \Tilde{\xi}_1 \leq \xi \leq 1 
    \end{cases}\label{eq:Qsine}
\end{equation}
\end{widetext}
\begin{equation}
    \langle \bar{Q}_1   \rangle=  1 - \cos (2\pi \tilde{\xi}_1)\label{eq:Qavesine}
\end{equation}

Two cases can be solved exactly: When $\kappa \rightarrow \infty$, $\Tilde{\xi}_1 \rightarrow 1$, and $\langle \bar{Q}_1 \rangle \rightarrow 0$; when $\kappa \rightarrow 0$, $\tilde{\xi}_1 \rightarrow \frac{1}{2}$, and $\langle \bar{Q}_1 \rangle \rightarrow 2$. For all finite $\kappa$, $\langle \bar{Q}_1 \rangle$ is positive.  
To quantify the effect a valve has on peristaltic flow, we review the leading-order results for small-amplitude peristalsis without valves. For a more complete study of valveless force-driven peristalsis, see \cite{Takagi_Balmforth_2011, Elbaz_Gat_2014}. The equations for the radius, pressure, and flow read:
\begin{equation}
    \bar{R}_1^{\text{nv}} (\xi)= \pi   \left[\frac{\sin (2\pi \xi) - \pi \kappa \cos (2\pi \xi)}{1+(\pi \kappa)^2}\right] ,\label{eq:RsineNV}
\end{equation}
\begin{equation}
    \bar{P}_1^{\text{nv}} (\xi)= \left[\frac{\cos (2\pi \xi) + \pi \kappa \sin (2\pi \xi)}{1+(\pi \kappa)^2}\right], \label{eq:PsineNV}
\end{equation}
\begin{equation}
    \bar{Q}_1^{\text{nv}} (\xi)= 2\pi   \left[\frac{\sin (2\pi \xi) - \pi \kappa \cos (2\pi \xi)}{1+(\pi \kappa)^2}\right], \label{eq:QsineNV}
\end{equation}
\begin{equation}
    \langle \bar{Q}_1^{\text{nv}} \rangle = 0, \hspace{.1cm }\langle \bar{Q}_2^{\text{nv}} \rangle = -4 \int_0^1 \bar{R}_1^{\text{nv}}(\xi') \frac{df}{d\xi}(\xi') d\xi' = \frac{4\pi^2}{1+(\pi \kappa)^{2}} .
\end{equation}
Note that the leading-order contribution to the mean flow is at order $\eta_P^2$, and the direction of the flow is always in the direction of peristalsis. We contrast these intuitive results with those predicted in the valve continuum where the leading-order contribution to the mean flow is order $\eta_P$, and we will soon find that the magnitude of the flow is independent of peristalsis direction, but always in the direction of the valve. The lowest-order term in the valveless problem is pure fluctuations that average to zero, but in the valve problem, these fluctuations are rectified. 

There is another interesting observation we can make by comparing the solutions with and without valves. The solutions in between open valves with zero valve resistance \eqref{eq:Rsine}, \eqref{eq:Psine}, \eqref{eq:Qsine} seem to closely resemble the solutions for the valveless problem \eqref{eq:RsineNV}, \eqref{eq:PsineNV}, \eqref{eq:QsineNV}. Other than some constant terms which enforce volume conservation, the only difference between the two solutions is a term proportional to $e^{-\frac{2}{\kappa} \xi}$. It is best to think of $\frac{\kappa}{2}$ as the diffusion coefficient as in equation \eqref{eq:diffusion}. Following a disturbance, the system relaxes on a timescale of $\frac{2}{\kappa}$. For a valveless system, this term is transient, but for a system with valves, this becomes part of the steady state. This term is particularly important when $\kappa$ is large such that the relaxation time is small compared to the period of peristaltic pumping. The extreme case $\kappa \rightarrow \infty $ corresponds to a system driven quasistatically, in which case the pressure diffuses completely such that $\bar{P}_1=0$, much like how a gas compressed quasistatically with a piston maintains a spatially uniform pressure throughout the process. 

It is difficult to continue our analysis to higher order in $\eta_P$ analytically due to the challenge in finding higher-order corrections to $\tilde{\xi}$, so even for moderate-amplitude peristalsis, one must resort to numerically solving \eqref{eq:continuityV}, \eqref{eq:momentumV}, and \eqref{eq:PR}. In the next couple subsections, we will further simplify the sinusoidal solutions to study the limits of large and small $\kappa$.

\subsubsection{Forward-Propagating, Small-Amplitude Peristalsis, Large $\kappa$}
Even at small amplitude, the condition for valve closure \eqref{eq: transcendental} is transcendental. We know that at $\kappa = \infty$, all but one valve is open, and $\Tilde{\xi}_1=1$. Perturbing $\Tilde{\xi}_1 \approx 1 - \delta \Tilde{\xi}_1$ and keeping the lowest order terms in $\delta \Tilde{\xi}_1$ and $\kappa^{-1}$ in \eqref{eq: transcendental} gives 
\begin{equation}
    \tilde{\xi}_{1} \approx 1-\frac{1}{\pi} \kappa^{-1/2} + \frac{1}{3\pi } \kappa^{-3/2} . \label{eq:xiTilde1_Stiff}
\end{equation}
The $\kappa^{-3/2}$ correction is only necessary for getting the correct pressure scaling. Even for large values of $\kappa$ (such as $\kappa=40$ in figure \ref{fig:stiffVC}), the fraction of valves open appears noticeably far from one due to the slowly varying $\kappa^{-1/2}$ term. The radius, pressure, flow, and mean flow can now easily be calculated:
\begin{equation*}
   \bar{R}_1(\xi)=
    \begin{cases}
       -\frac{1}{\kappa} \cos(2\pi \xi),& \text{for } 0\leq \xi \leq \Tilde{\xi}_1 \\
       -\frac{1}{ \kappa},& \text{for } \Tilde{\xi}_1 \leq \xi \leq 1, 
    \end{cases}
\end{equation*}
\begin{equation*}
   \bar{P}_1(\xi)=
    \begin{cases}
       -\frac{2}{\kappa}\left[(\xi - \frac{1}{2})-\frac{1}{2\pi } \sin (2\pi \xi)\right],& \text{for } 0\leq \xi \leq \Tilde{\xi}_1 \\
       -1+\cos(2\pi \xi)+\frac{1}{ \kappa},& \text{for } \Tilde{\xi}_1 \leq \xi \leq 1, 
    \end{cases}
\end{equation*}
\begin{equation*}
   \bar{Q}_1(\xi) =
    \begin{cases}
      \frac{2}{\kappa}\left[1- \cos (2\pi \xi) \right],& \text{for } 0\leq \xi \leq \Tilde{\xi}_1 \\
      0,& \text{for } \Tilde{\xi}_1 \leq \xi \leq 1 ,
    \end{cases}
\end{equation*}
\begin{equation*}
    \langle \bar{Q}_1   \rangle= \frac{2}{\kappa}.
\end{equation*}
These solutions for $\Bar{R}_1$ and $\bar{Q}_1$ agree with the $O(\eta_R)$ solutions in section \ref{sec:stiffFW}, but now we also have estimates for the fraction of open valves and the pressure in the region of closed valves. The pressure becomes discontinuous for infinite stiffness, but we now see that for any finite stiffness, the pressure in the small region of closed valves continuously interpolates to the solutions in the open regions. Although we were able to explain the mean flow in section \ref{sec:stiffFW} purely as a result of the continuity equation (and constraints placed by the valves), we can now also understand the mean flow as a consequence of the linear pressure drop established across the open region of valves. Physically, this pressure drop arises from the rectification of the diffusive motion described by \eqref{eq:diffusion} for the valveless case. 

\subsubsection{Forward-Propagating, Small-Amplitude Peristalsis, Small $\kappa$}
At small $\kappa$, we can neglect the exponential term in \eqref{eq: transcendental}. We will keep the exponential term in \eqref{eq:Rsine}, \eqref{eq:Psine}, and \eqref{eq:Qsine} to ensure the boundary condition at $\xi=0$ is satisfied, and write \eqref{eq: transcendental}, \eqref{eq:Rsine}, \eqref{eq:Psine}, \eqref{eq:Qsine} and \eqref{eq:Qavesine} as:
\begin{equation*}
    \tilde{\xi} = \frac{1}{2} + \frac{\kappa}{2},
\end{equation*}
\begin{widetext}
\begin{equation*}
   \bar{R}_1(\xi)=
    \begin{cases}
       -1 + \pi \sin(2\pi \xi)+\pi^2 \kappa \left(e^{-\frac{2}{\kappa}\xi} -\cos(2\pi \xi) \right),& \text{for } 0\leq \xi \leq \Tilde{\xi}_1 \\
      -1,& \text{for } \Tilde{\xi}_1 \leq \xi \leq 1 ,
    \end{cases}
\end{equation*}
\begin{equation*}
    \bar{P}_1(\xi)=
    \begin{cases}
        \cos(2\pi \xi) -\kappa + \pi \kappa \sin(2\pi \xi) + \pi^2 \kappa^2 e^{-\frac{2}{\kappa} \xi },& \text{for } 0\leq \xi \leq \Tilde{\xi}_1 \\
        \cos (2\pi \xi)-\kappa,& \text{for } \Tilde{\xi}_1 \leq \xi \leq 1 ,
    \end{cases}
\end{equation*}
\begin{equation*}
   \bar{Q}_1(\xi) =
    \begin{cases}
     2\pi \sin(2\pi \xi)+2\pi^2 \kappa \left(e^{-\frac{2}{\kappa}\xi} -\cos(2\pi \xi) \right),& \text{for } 0\leq \xi \leq \Tilde{\xi}_1 \\
      0,& \text{for } \Tilde{\xi}_1 \leq \xi \leq 1 ,
    \end{cases}
\end{equation*}
\end{widetext}
\begin{equation}
    \langle \bar{Q}_1 \rangle = 2. \label{eq:Q1_compliant}
\end{equation}
The peristaltic period is much shorter than the elastic response time, so the fluid pressure is dominated by the external pressure, and the radial deformation is out of phase with $f$. Since $f$ has negative slope half of the time, $\bar{P}_1$ also has negative slope about half of the time, suggesting about half of the valves should be open. Up to some constants fixed by matching conditions and the exponentially small term, the solutions in the open region exactly match those for the valveless problem. The diffusion mechanism discussed in the previous subsection is suppressed. 

An example solution in this regime is given in the left column of figure \ref{fig:convergence}$c$. Because the vessel is highly compliant, even though $\eta_P$ is small, the radial deformation is large. The pressure for the cases with and without valves is similar, so the flow with valves is just the rectification of the flow without valves: 
\begin{equation*}
    \bar{Q}_1 \approx  \bar{Q}_1^{\text{nv}} \Theta (\bar{Q}_1^{\text{nv}}).
\end{equation*}

\subsection{Forward-Propagating, Large-Amplitude Peristalsis}
Large-amplitude peristalsis is already capable of pushing fluid in one direction, so if the peristalsis direction agrees with the valve direction, the flow response of the valve problem will essentially be that of the valveless problem. This is confirmed in the left column of figure \ref{fig:convergence}$d$. The radius is narrow throughout most of the tube, but is very large in a small region which traps and pushes the fluid in the direction of peristalsis. Since the fluid is completely trapped, the mean flow approaches one in the limit $\eta_P \rightarrow \infty$. The deviation from one can be found by considering the fluid in the occluded region where both $\bar{R}^{-4}$ and $\eta_P$ are large such that the $\Bar{R}'$ term in equation \ref{eq:Rvc} is negligible. Then $\Bar{R}^2$ satisfies an algebraic equation whose solution is real if and only if $\Bar{R}^2(0) \leq \frac{1}{8\pi \eta_P}$. This leads to the scaling 
\begin{equation}
    1-\langle \bar{Q} \rangle \approx \frac{1}{8\pi \eta_P}.
\end{equation}
A thorough analysis of large-amplitude force-imposed peristalsis can be found in \cite{Takagi_Balmforth_2011}. Although it is mathematically unsurprising that the valve system behaves like a valveless system when driven by large-amplitude peristalsis, it is worth stressing the implication of this. If a biological or engineered system is reliably driven by large-amplitude peristalsis, there is no reason for it to contain valves. We have already seen one reason why a system may utilize valves. Given a small perturbation, the valveful system can transport volume proportional to the amplitude of the perturbation (as opposed to the amplitude squared as is the case for valveless peristalsis). We will see an even more apparent difference when we consider backward-propagating peristaltic waves in the next section. 

\begin{figure*}
    \centering
     \includegraphics[width=.85\textwidth]{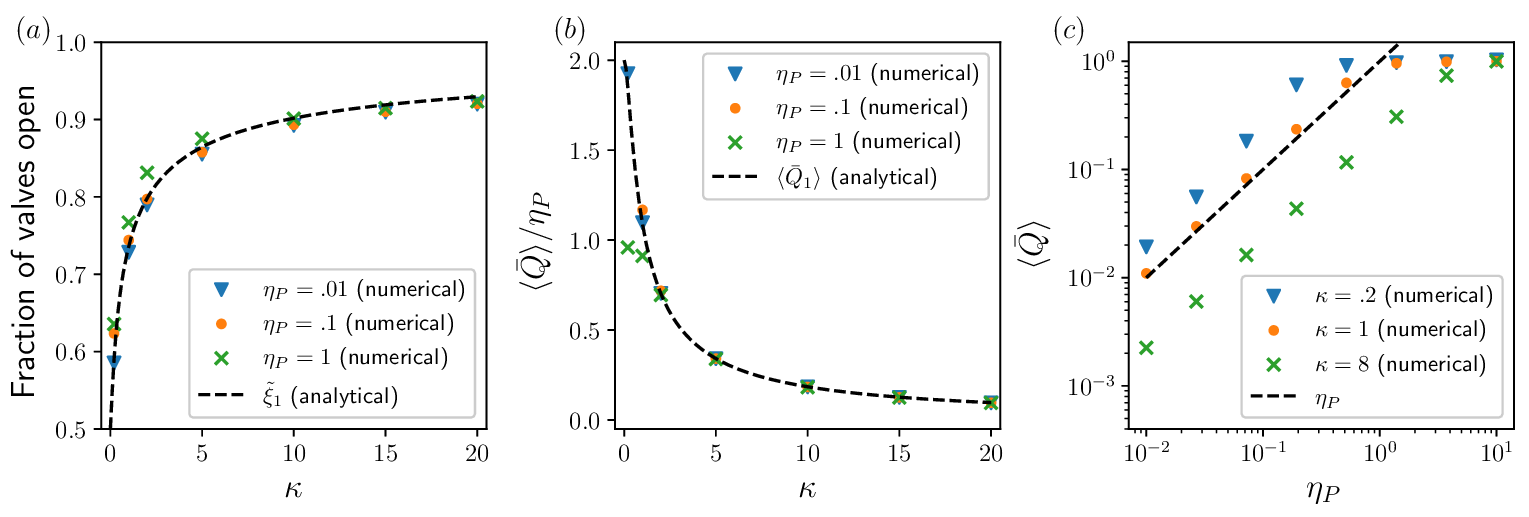}
    \caption{Results for forward-propagating peristalsis with a continuum of valves, $f(\bar{x},\Bar{t}) = \cos(2\pi (\Bar{x}-\Bar{t}))$. $(a)$ The fraction of valves per wavelength which are open at any given time is calculated numerically (points) as a function of $\kappa$ and compared with the small-amplitude result \eqref{eq: transcendental} (dashed line). $(b)$ The mean flow divided by $\eta_P$ is calculated numerically as a function of $\kappa$ and compared with the small-amplitude result \eqref{eq:Qavesine}. $(c)$ The mean flow is calculated as a function of $\eta_P$. For small $\eta_P$, the scaling is linear for each $\kappa$, as demonstrated by the dashed line. }
    \label{fig:xiQbarFW}
\end{figure*}
 
A complete summary of the fraction of valves open and the mean flow for the case of a forward-propagating sinusoidal wave is shown in figure \ref{fig:xiQbarFW}. The fraction of valves open is well-approximated by the small-amplitude result even for $\eta_P=1$, as shown in figure \ref{fig:xiQbarFW}$a$. Though, for larger values of $\eta_P$, this is a less useful metric for quantifying the system since the behavior is nearly that of the valveless system. In figure \ref{fig:xiQbarFW}$b$, it is shown that for a fixed $\eta_P$, the flow decreases with $\kappa$, indicating a larger flow response in more compliant tubes subject to forces of the same magnitude. In figure \ref{fig:xiQbarFW}$c$, it is clear that the flow for small $\eta_P$ is linear in $\eta_P$, a unique feature to the system with ideal valves, but for $\eta_P$ close to or greater than one, the flow approaches the fully occluded limit $\langle \bar{Q} \rangle = 1$.

\section{Valve Continuum Results II: Backward-Propagating Peristaltic Waves} \label{sec:valve_continuumBW}
In this section, we will revisit some of the calculations from the previous section but with backward-propagating peristaltic forces of the form 
\begin{equation}
    f(\Bar{x},\Bar{t}) = f(\Bar{x}+\Bar{t}) \equiv f(\xi).
\end{equation}
The origin is again chosen such that closing occurs at $\xi = 0$ and opening occurs at $\xi = \Tilde{\xi}$. Equation \eqref{eq:momentumV} is unchanged, but the continuity equation \eqref{eq:continuityV} becomes
\begin{equation}
    \frac{d }{d \xi} \Big[ \bar{Q} + \bar{R}^2 \Big] = 0 \implies \Bar{Q}(\xi) =  \Bar{R}^2(0) - \Bar{R}^2(\xi)\label{eq:continuityVBW} .
\end{equation}
Following similar arguments as in section \ref{sec:valve_continuumFW}, $\Bar{R}(0) = \max \Bar{R}$ and closing occurs at $\min \Bar{P}$. Therefore, 
\begin{equation}
    f(0) = \min f \iff \text{valve closes at $\xi = 0$}.
\end{equation}
An illustration is given in figure \ref{fig:demo}$b$. The decoupled radius equation is 
\begin{equation}
    \begin{cases}
        \Bar{R}(\xi) = \Bar{R}(0) & 0\leq \xi \leq \Tilde{\xi} \\
        \frac{d\bar{R}}{d\xi} =  \frac{1}{\kappa} \bar{R}^{-4} \Big(\bar{R}^{2} - \bar{R}(0)^2  \Big) - \frac{\eta_P}{\kappa} \frac{df}{d\xi}  & \Tilde{\xi} \leq \xi \leq 1, 
    \end{cases} \label{eq:dRdxiBW}
\end{equation}
where $\Tilde{\xi} \in (0,1]$ is defined by matching the solutions, and the constant $\Bar{R}(0)$ is fixed by enforcing volume conservation \eqref{eq:RSquaredConstraint} which in this case takes the special form
\begin{equation}
    \tilde{\xi} \Bar{R}^2(0) 
 + \int_{\Tilde{\xi} }^1\Bar{R}^2(\xi) d\xi = 1.  \label{eq:area_constraintBW}
\end{equation}
Note that the fraction of time open for backward-propagating peristalsis as we have formulated it is $1-\Tilde{\xi}$. The mean flow is 
\begin{align}
     \langle \bar{Q} \rangle = \Bar{R}^2(0) - 1 . \label{eq:QbarVC_BW}
\end{align}
The fact that $\Bar{R}(0) = \max \Bar{R}$ automatically suggests something peculiar. Typically, we think of the flow driven by peristalsis as being largest in regions where the tube has expanded, but here, the radius is largest in regions where the valves are closed, and the flow is zero. In regions where the valves are open, and flow is permitted, $\Bar{R}$ will be smaller than one, and the nonlinear factor of $\Bar{R}^4$ in the momentum equation will suppress the flow. See the right columns of figure \ref{fig:convergence} for some example solutions. 

The backward-propagating wave has an additional constraint. Because $\langle \Bar{Q} \rangle $ must remain less than 1 (corresponding to the surprising case where all fluid volume is transported forward each period), \eqref{eq:QbarVC_BW} implies that $\Bar{R}(0)^2<2$, and since $\bar
R(0)=\max \Bar{R}$, we have more generally that
\begin{equation}
    \Bar{R}(\xi) \leq \sqrt{2}. \label{eq:BWConstraint}
\end{equation}
The forward-propagating peristaltic wave possessed no such bound on the radius for either the valveless or the valve continuum problem, but this constraint will be important for understanding large-amplitude retrograde peristalsis with many valves.

\subsection{Backward-Propagating Peristalsis in a Stiff Tube (Radius-imposed peristalsis)}\label{sec:stiffBW}
 In the limit $\kappa \rightarrow \infty$ at fixed small $\eta_R$, we again recover radius-imposed peristalsis with $\bar{R}$ given by \eqref{eq:RstiffFW}, but in this case, the flow is given by 
\begin{align}
    \bar{Q}(\xi) &=  2 \eta_R \sqrt{1-\eta_R^2 \langle f^2 \rangle} \Big(f(\xi) -\min (f) \Big) \nonumber \\ &\hspace{.4cm}-\eta_R^2 \left(f(\xi)^2 - (\min (f) )^2\right) \label{eq:QStiffBW}
\end{align}
\begin{align}
    \langle \Bar{Q} \rangle &=  -2 \eta_R \min (f)  \sqrt{1-\eta_R^2 \langle f^2 \rangle} \nonumber \\ &\hspace{.4cm} + \eta_R^2 \left((\min (f) )^2 - \langle f^2 \rangle \right) .\label{eq:QbarStiffBW}
\end{align}
This looks similar to the forward peristalsis solution. In fact, \eqref{eq:QbarStiffBW} suggests that for a sine wave, the flow induced by reverse peristalsis is larger than that induced by forward peristalsis satisfying \eqref{eq:QbarStiffFW}. The caveat is that \eqref{eq:QbarStiffBW} is valid over a smaller range of values than \eqref{eq:QbarStiffFW}. To see this, note that in addition to satisfying \eqref{eq:etaRconstraint}, we also need to satisfy \eqref{eq:BWConstraint} which gives an additional constraint 
\begin{equation}
    \eta_R < \frac{1}{\sqrt{2}|\min f|+\sqrt{(\min f)^2 - \langle f^2 \rangle }}. \label{eq:etaRconstraint2}
\end{equation}
For the case of a sine wave, \eqref{eq:etaRconstraint} gives a bound of $\eta_R<\sqrt{2/3}$, but \eqref{eq:etaRconstraint2} places a stronger bound of $\eta_R < \sqrt{2/9}$. To be clear, the bound only tells us when the analytic solution is guaranteed to give an unphysical solution, it does not tell us when the radius-imposed solutions will work. Although we found good agreement with the analytic result for all physical values in the case of a forward-propagating wave in a stiff tube, the same is not true here. In fact, it appears the analytic solution \eqref{eq:QStiffBW} only works for values much smaller than \eqref{eq:etaRconstraint2}. In that limit, only the linear term in \eqref{eq:QbarStiffBW} contributes, and for the case of a sine wave, the mean flow is identical to that of a forward-propagating wave. Indeed, comparing the left and right columns of figures \ref{fig:convergence}$a$ and $b$, the radius and flow look nearly identical. For an $f$ with minimum value larger in magnitude than its maximum value, the small-amplitude backward-propagating wave will produce more forward flow than the small-amplitude forward-propagating wave. This effect is demonstrated using gaussian wave forms in the appendix.

\begin{figure*}
    \centering
     \includegraphics[width=.85\textwidth]{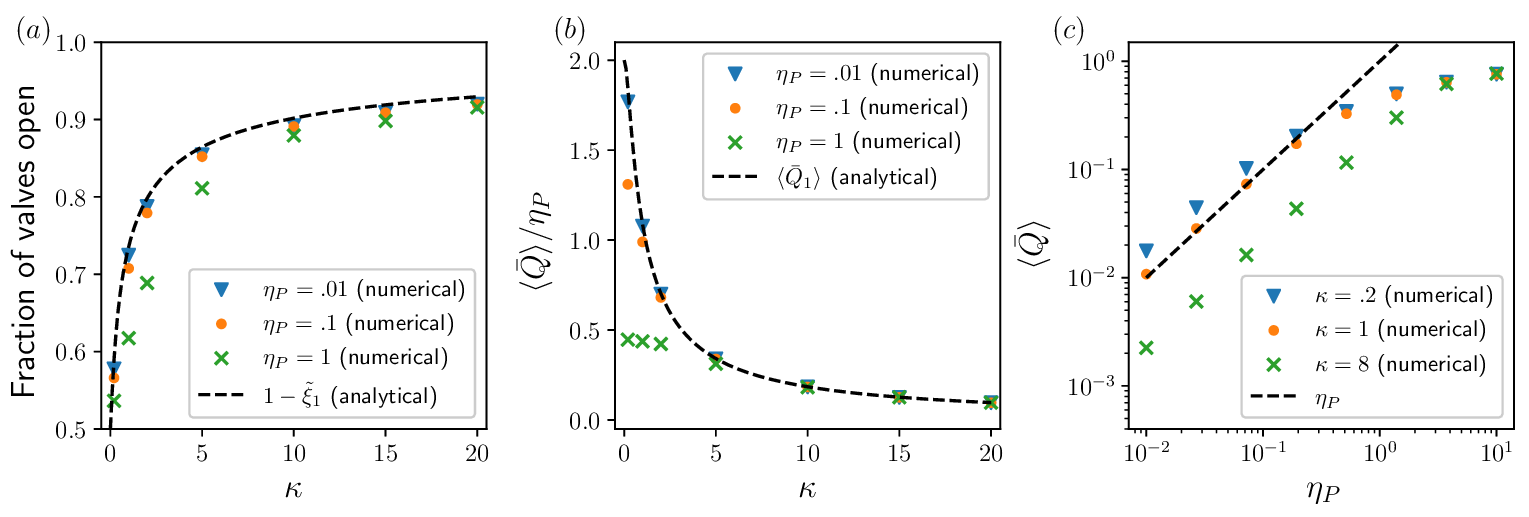}
    \caption{Results for backward-propagating peristalsis with a continuum of valves, $f(\bar{x},\Bar{t}) = -\cos(2\pi (\Bar{x}+\Bar{t}))$. The dashed lines in $(a)$ and $(b)$ correspond to the small-amplitude analytic expressions \eqref{eq:transcendentalBW} and \eqref{eq:Qbar1BW}, which are identical to equations \eqref{eq: transcendental} and \eqref{eq:Qavesine} for the forward-propagating wave. }
    \label{fig:xiQbarBW}
\end{figure*}

\subsection{Backward-Propagating, Small Amplitude Peristalsis}
Proceeding as in section \ref{sec:Small_Amplitude_FW}, we wish to solve
\begin{equation}
    \begin{cases}
         \Bar{R}_1(\xi) = \Bar{R}_1(0) & 0\leq \xi \leq \Tilde{\xi}_1 \\
        \frac{d\bar{R}_1}{d\xi} =  \frac{2}{\kappa} \left(\bar{R}_1-\bar{R}_1(0)\right) - \frac{1}{\kappa}\frac{df}{d\xi}  & \Tilde{\xi}_1 \leq \xi \leq 1.
    \end{cases} 
\end{equation}
In this case, $\Tilde{\xi}_1$ satisfies satisfies
\begin{equation}
   \int_{\Tilde{\xi}_1}^1 d\xi' \frac{df(\xi')}{d\xi} e^{-\frac{2}{\kappa} (\xi'-\Tilde{\xi}_1)} = 0,
\end{equation}
and the remaining quantities of interest are
\begin{widetext}
\begin{equation}
    \bar{R}_1(\xi)=
    \begin{cases}
     \frac{1}{2}[f(\tilde{\xi}_1)-f(0)] & \text{for } 0\leq \xi \leq \Tilde{\xi}_1 \\
      \frac{1}{2}[f(\tilde{\xi}_1)-f(0)] - \frac{1}{\kappa} \int_{\Tilde{\xi}_1}^{\xi} d\xi' \frac{df(\xi')}{d\xi} e^{-\frac{2}{\kappa} (\xi'-\xi)} & \text{for } \Tilde{\xi}_1\leq \xi \leq 1 ,
    \end{cases} 
\end{equation}
\begin{equation}
    \bar{P}_1(\xi)=
    \begin{cases}
     \frac{\kappa}{2}[f(\tilde{\xi}_1)-f(0)] +f(\xi) & \text{for } 0\leq \xi \leq \Tilde{\xi}_1 \\
      \frac{\kappa}{2}[f(\tilde{\xi}_1)-f(0)] +f(\xi) - \int_{\Tilde{\xi}_1}^{\xi} d\xi' \frac{df(\xi')}{d\xi} e^{-\frac{2}{\kappa} (\xi'-\xi)} & \text{for } \Tilde{\xi}_1\leq \xi \leq 1 ,
    \end{cases} 
\end{equation}
\begin{equation}
    \bar{Q}_1(\xi)=
    \begin{cases}
      0& \text{for } 0\leq \xi \leq \Tilde{\xi}_1 \\
       \frac{2 }{\kappa}  \int_{\tilde{\xi}_1}^\xi d\xi' \frac{df}{d\xi} e^{-\frac{2}{\kappa}(\xi'-\xi)}& \text{for } \Tilde{\xi}_1\leq \xi \leq 1 ,
    \end{cases} 
\end{equation}
\end{widetext}
\begin{align}
    \langle \bar{Q}_1 \rangle &= 2 \bar{R}_1(0) = f(\tilde{\xi}_1)-f(0) .
\end{align}
For the special case of a sine wave $f(\xi) = -\cos(2\pi \xi)$, the fraction of time open satisfies the same equation as that of the forward-propagating wave:
\begin{equation}
    e^{-\frac{2}{\kappa} (1-\Tilde{\xi}_1)} - \cos \left(2\pi (1-\Tilde{\xi}_1) \right)+ \frac{1}{\pi \kappa} \sin \left(2\pi (1-\Tilde{\xi}_1)\right) = 0 . \label{eq:transcendentalBW}
\end{equation}
The mean flow is also identical:
\begin{equation}
    \langle \bar{Q}_1   \rangle =  1 - \cos \left(2\pi (1-\tilde{\xi}_1)\right).\label{eq:Qbar1BW}
\end{equation}
Figures \ref{fig:xiQbarBW}$a$ and $b$ confirm this claim numerically. For $\eta_P\ll 1$, the numerical results agree with the predictions for $1-\Tilde{\xi}_1$ and $\langle \Bar{Q}_1\rangle$. It is not, in general, true that an arbitrary low-amplitude peristaltic wave in a valve-filled tube produces the same flow regardless of pumping direction.

\begin{figure*}
    \centering
    \includegraphics[width=.85\textwidth]{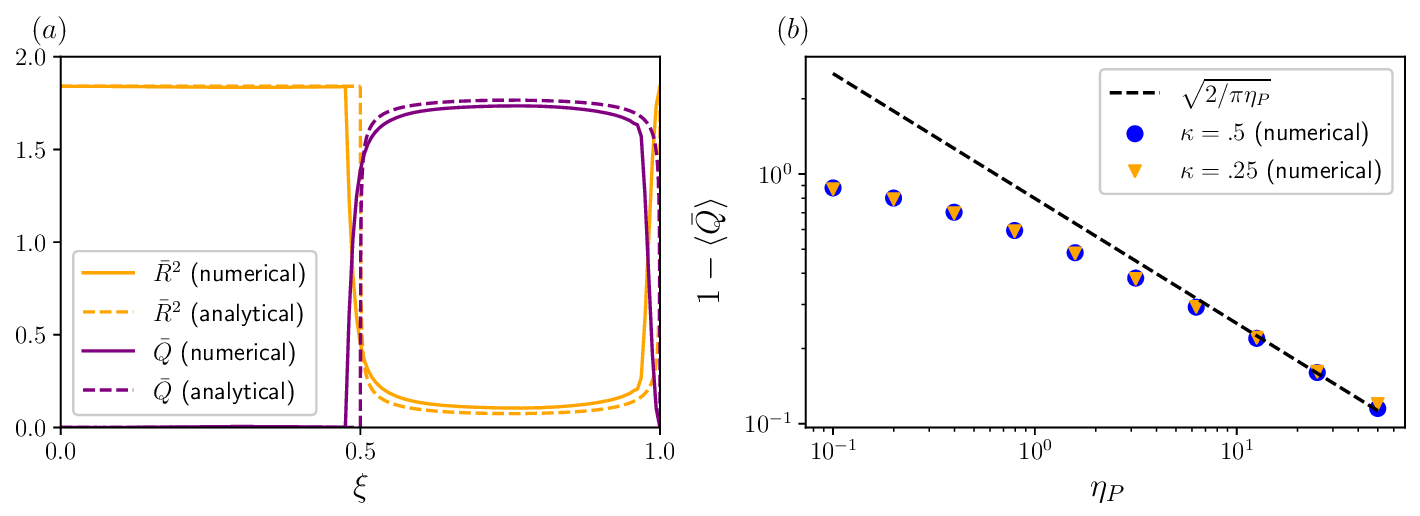}
    \caption{Large-amplitude peristalsis against a continuum of valves with $f(\bar{x}, \bar{t}) = - \cos(2\pi (\bar{x}+\bar{t}))$. $(a)$ Numerical solutions for the cross-sectional area and flow using $\eta_P = 25, \kappa = .5$ are shown with solid lines and compared with the analytic solution \eqref{eq:RSquaredLargeAmplitudeBW} with $\Bar{R}(0)^2$ given by \eqref{eq:maxRSq} shown with dotted lines. $(b)$ Deviation of the mean flow from the fully occluded limit. The dots show numerical solutions for two different choices of $\kappa$, and the dashed line is the analytical large-amplitude prediction \eqref{eq:OneMinusQ}.}
    \label{fig:LargeAmplitudeBW}
\end{figure*}

\subsection{Backward-Propagating, Large-Amplitude Peristalsis}
For the amplitudes considered in figure \ref{fig:xiQbarBW}$c$, the flow appears to reach 1 at a slower rate as compared with figure \ref{fig:xiQbarFW}$c$. Unlike forward-propagating peristaltic waves, large-amplitude backward-propagating peristaltic waves produce solutions which are qualitatively different from the valveless problem. For backward-propagating peristaltic waves, the tube becomes constricted in the region of open valves, and the radius approaches the maximum value \eqref{eq:BWConstraint} in the closed regions. For $\kappa$ not too large, we expect the terms on the right side of \eqref{eq:dRdxiBW} to dominate since $\Bar{R}^{-4}$ and $\eta_P$ are both large. Setting $R'(\xi)=0$, we are left solving an algebraic equation which is independent of $\kappa$:
\begin{equation}
    - \eta_P \frac{df}{d\xi} \Bar{R}^4 + \bar{R}^{2} - \bar{R}(0)^2 = 0 .
\end{equation}
Just like the $\kappa \rightarrow 0$ limit, the solution to this equation is to have $f'(\Tilde{\xi}) = 0$, so the open region of valves will correspond to where $f$ is decreasing. With this in mind, we can uniquely pick the sign to this quadratic equation that gives a positive area: 
\begin{equation}
  \bar{R}(\xi)^2 =  \frac{-1+ \sqrt{1+4\Bar{R}(0)^2 \eta_P \big|f'(\xi)\big|}}{2\eta_P \big|f'(\xi)\big|}. \label{eq:RSquaredLargeAmplitudeBW}
\end{equation}
In order to fix $\Bar{R}(0)$, we would need to solve \eqref{eq:area_constraintBW}. This is difficult to solve in general, but we can approximately solve this problem for the case of a sine wave where $\Tilde{\xi}=\frac{1}{2}$. We will assume that the integral is dominated by the largest values of the radius close to the closed regions, where $\sin(2\pi \xi)$ is approximately linear.
\begin{widetext}
\begin{align*}
    1 &= \frac{\Bar{R}(0)^2}{2} + \int_{\frac{1}{2}}^1 \frac{-1+ \sqrt{1+8\pi \Bar{R}(0)^2 \eta_P |\sin (2\pi \xi)|}}{4\pi \eta_P |\sin (2\pi \xi)|} d\xi \\
    &= \frac{\Bar{R}(0)^2}{2} + 2 \int_{0}^{\frac{1}{4}} \frac{-1+ \sqrt{1+8\pi \Bar{R}(0)^2 \eta_P \sin (2\pi \xi)}}{4\pi \eta_P \sin (2\pi \xi)} d\xi \\
    &\approx \frac{\Bar{R}(0)^2}{2} + 2 \int_{0}^{\frac{1}{4}} \frac{-1+ \sqrt{1+8\pi \Bar{R}(0)^2 \eta_P  (2\pi \xi)}}{4\pi \eta_P (2\pi \xi)} d\xi \\
    &= \frac{\Bar{R}(0)^2}{2} + \frac{1}{\pi \eta_P} \left[ \sqrt{1+2\pi \Bar{R}(0)^2 \eta_P } - 1 - \log \left(\frac{1+\sqrt{1+2\pi \Bar{R}(0)^2 \eta_P}}{2} \right) \right] \\
    &= \frac{\Bar{R}(0)^2}{2} + \sqrt{\frac{2}{\pi \eta_P}} \bar{R}(0) + O\left( \frac{\log \eta_P}{\eta_P}\right)
\end{align*}
\end{widetext}
We can now write down the maximum radius and mean flow:
\begin{equation}
    \implies (\max \Bar{R})^2\approx 2 - \sqrt{\frac{2}{\pi \eta_P}} \label{eq:maxRSq}
\end{equation}
\begin{equation}
    1 - \langle \Bar{Q} \rangle \approx \sqrt{\frac{2}{\pi \eta_P}}. \label{eq:OneMinusQ}
\end{equation}
This is a distinct power law from the forward case which approaches 1 as $\eta_P^{-1}$. It is not surprising that much larger amplitudes are needed in order to achieve maximum flow for a backward-propagating wave as compared to a forward-propagating wave. Perhaps it is more surprising that this limit is ever achieved. As a technical note, one can also derive an $\eta_P^{-1/2}$ power law without making the approximation on the third line but instead assuming that the $4\Bar{R}(0)^2 \eta_P f'(\xi)$ term dominates the square root; however, this gives a slightly smaller prefactor which does not agree as well with our numerical results. The analytical results are confirmed in figure \ref{fig:LargeAmplitudeBW}. Although the radius is small in the open region, the large imposed force produces a pressure gradient which drives flow. This is entirely different from large-amplitude peristalsis without valves where the flow is confined to a region of large radius but small pressure gradient.

\section{Valve density}\label{sec:density}
\begin{figure*}
    \centering
    \includegraphics[width=.85\textwidth]{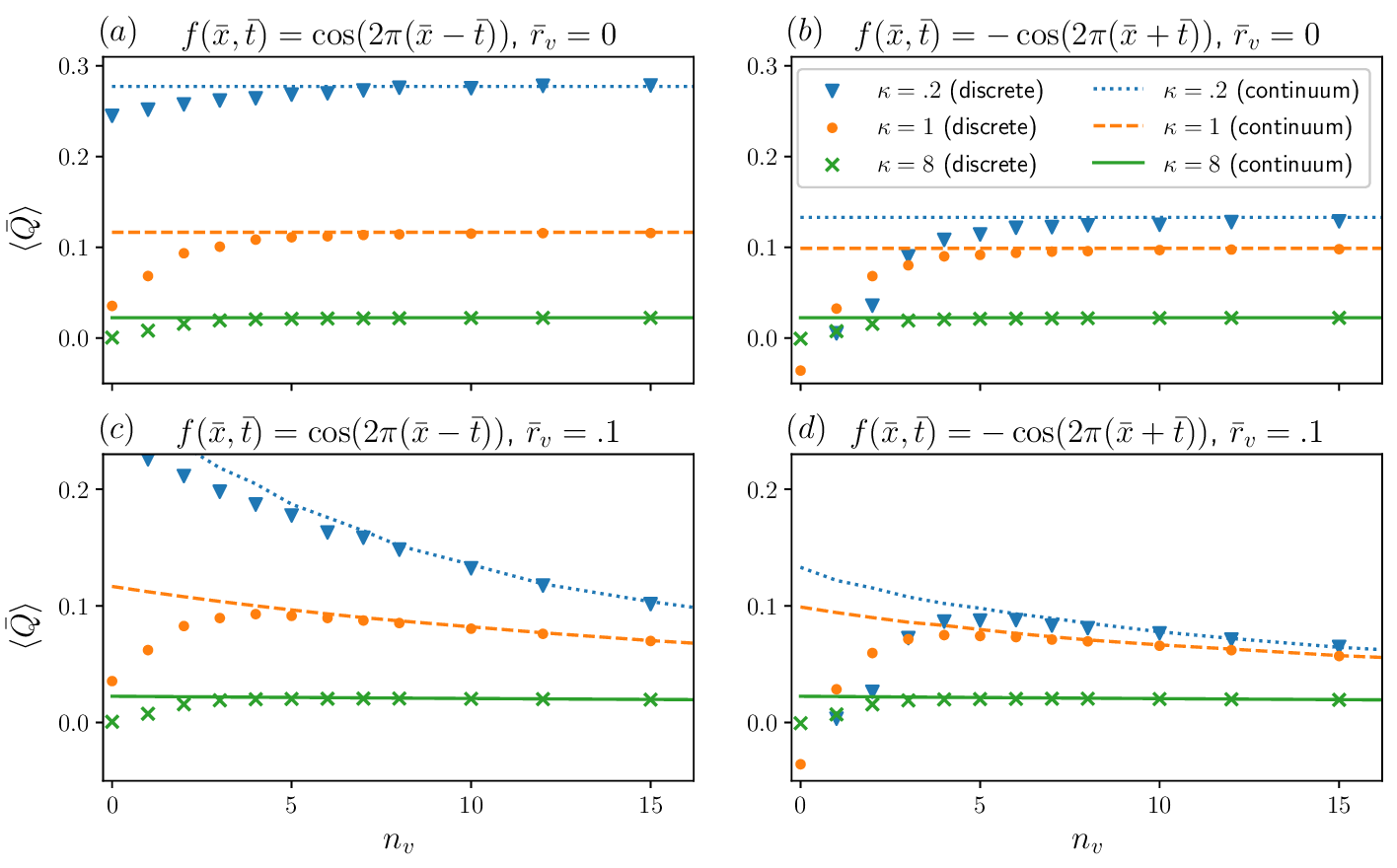}
    \caption{Mean flow as a function of the number of equally spaced valves per wavelength $n_v $. As the density of valves increases, the mean flow approaches the valve continuum result (lines). In all cases, $\eta_P=0.1$. $(a)$ Forward-propagating peristaltic wave with zero valve resistance. $(b)$ Backward-propagating peristaltic wave with zero valve resistance. $(c)$ Forward-propagating peristaltic wave with nonzero valve resistance. $(d)$ Backward-propagating peristaltic wave with nonzero valve resistance. }
    \label{fig:rv}
\end{figure*}

The previous sections demonstrated the features of the valve continuum model with $\Bar{r}_v=0$. Valves are able to successfully pump fluid given only small-amplitude perturbations or to pump fluid against the direction of peristalsis. In any case, if $\Bar{r}_v=0$, the mean flow is optimized by having a continuum of valves, as demonstrated in figure \ref{fig:rv}$a,b$. Notice that for a system with more than about 5 valves, the valve continuum prediction of the mean flow is nearly correct. However, when $\Bar{r}_v>0$, an excessive number of valves will increase the fluidic resistance and thus lower the magnitude of the flow, as demonstrated in \ref{fig:rv}$c,d$. The valve continuum prediction, in this case, is numerically obtained by filling all edges with valves of zero resistance, but with modified parameters $\kappa \rightarrow \kappa/(1+n_v \Bar{r}_v)$ and $\eta_P \rightarrow \eta_P/(1+n_v \Bar{r}_v)$. It appears that for a stiff vessel ($\kappa=8$ in figure \ref{fig:rv}), the flow barely changes as more valves are added. This is because, as we saw in sections \ref{sec:stiffFW} and \ref{sec:stiffBW}, the flow in this limit only depends on $\eta_R$ which is not modified by having a nonzero $\Bar{r}_v$. All but one valve is kept open regardless of the number of valves, and the result is completely determined by the continuity equation, so it is not too surprising that the number of valves is unimportant in this regime. For small $\kappa$ and small $\eta_P$, equation \eqref{eq:Q1_compliant} suggests that the mean flow will decay with the number of valves as $n_v^{-1}$. Comparing the triangles in figures \ref{fig:rv}$c$ and $d$, it is clear that valves may sometimes be unnecessary in rectifying flow when peristalsis is driven by a forward-propagating wave, but necessary when peristalsis is driven by a backward-propagating wave. 

\begin{figure*}
    \centering
    \includegraphics[width=.85\textwidth]{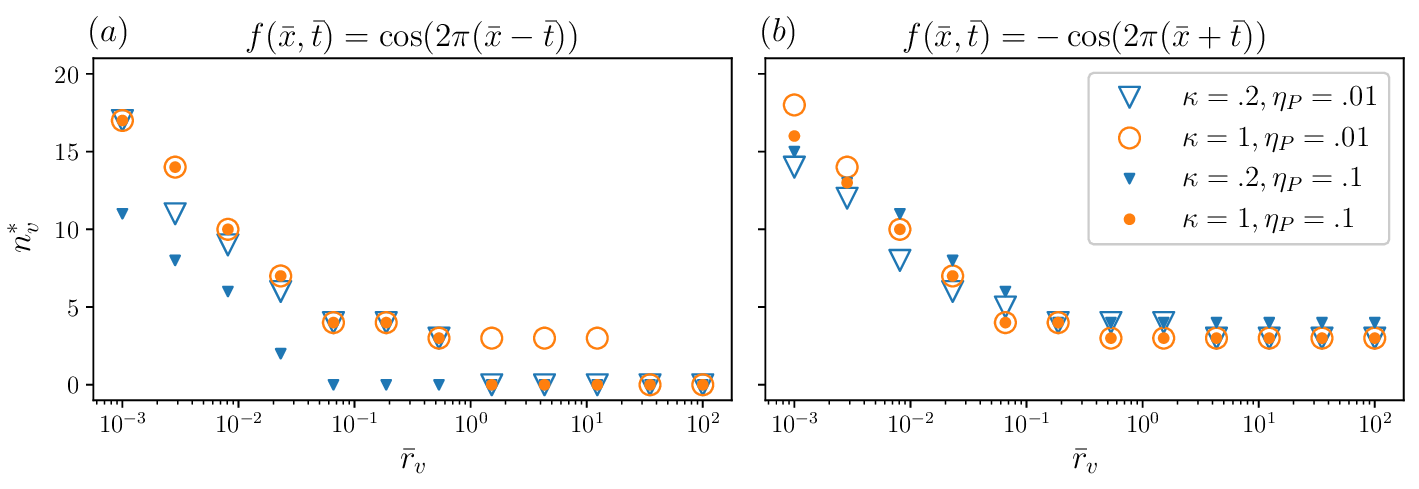}
    \caption{The value of $n_v$ which maximizes $\langle \bar{Q} \rangle$, denoted $n_v^*$, as a function of valve resistance $\Bar{r}_v$. Valves were spaced equally, and all integer values of $n_v$ between 0 and 25 were tested. $(a)$ Forward-propagating peristaltic wave. $(b)$ Backward-propagating peristaltic wave.}
\label{fig:nvStar}
\end{figure*}

One enlightening application of this model is to estimate the density of valves that optimizes the mean flow (see e.g. \cite{Venugopal2009a}). The most interesting case is when $\eta_P$ is small and $\bar{r}_v \neq 0$. In this case, there is a finite optimum number of valves that maximizes the mean flow which will be denoted $n_v^*$. We will avoid studying the case of large $\kappa$ since there is only a slight dependence on $n_v$ in this limit. The results are shown in figure \ref{fig:nvStar}. In panel $(a)$, it is demonstrated that for forward-propagating peristaltic waves, $n_v^*>0$ when $\bar{r}_v$ is small, but $n_v^*=0$ when $\Bar{r}_v$ is large. Also, $n_v^*$ is smaller when $\eta_P$ is larger since valveless peristalsis is more efficient in that case. Contrast these results with those in panel $(b)$ where the fluid is driven by backward-propagating peristaltic waves. Here, there is a less clear dependence on $\eta_P$, and even for very large values of $\Bar{r}_v$, we find $n_v^*>0$. Interestingly, for several orders of magnitude, having 3 or 4 valves per wavelength seems optimal when $\Bar{r}_v$ is large.

\section{Discussion} \label{sec:discussion}
\subsection{Application: Lymphatic system}
In order to apply the results of this paper to the lymphatic system, we must first check that our models for the fluid, vessel, and valve are appropriate. If so, then we may check that approximation \eqref{eq:approx} holds and study the implications of our valve continuum model. 

The viscosity of lymph is nearly that of water \citep{Macdonald_Arkill_Tabor_McHale_Winlove_2008}, but due to the small radius of lymphangions, the fluid is often in the Stokes regime. Lymph can be treated at zero Reynolds number when the diameters are less than 100 $\mu$m \citep{Moore_Bertram_2018}. This is the case for the rat mesenteric lymphatics, where the Reynolds number was experimentally calculated to be .045  \citep{Dixon_Greiner_Gashev_Cote_Moore_Zawieja_2006}. However, in the largest lymphatic vessel in the human body, the thoracic duct, the diameters are closer to 2 mm, and inertial effects should be taken into account \citep{Moore_Bertram_2018}. 

Experimental data on lymphatic pumping is plentiful, but few studies have been able to resolve the propagation of peristaltic waves. When only a single lymphangion is studied, the whole vessel appears to contract uniformly \citep{Moore_Bertram_2018}. Thus, the ratio $x_v/\lambda$ is difficult to find in the literature, while $R_0/x_v$ can be found in numerous sources. A detailed study of the peristaltic waves in mesenteric rat lymphatics is given in \cite{Zawieja_Davis_Schuster_Hinds_Granger_1993}. The authors demonstrate that 80-90 percent of waves are coordinated, meaning contractions are observed in adjacent sites within one second of each other. The waves were observed to propagate at a speed of $4 - 8$ mm/s at a contraction frequency of $8.4 - 13$/min. The calculated wavelength is $\lambda = 20-60$ mm. The radii of these vessels was measured to be $R_0=.03 - .06$ mm, and $x_v = .6 - 1$ mm, so there are about 20-100 valves per wavelength. Thus, we can safely say $R_0\ll x_v \ll \lambda$. For bovine mesenteric lymphatics, the contraction waves were seen to propagate at $4-5$ mm/s  with a frequency of $4-6$/min, indicating a wavelength of $\lambda = 40 - 75$ mm \citep{Ohhashi_Azuma_Sakaguchi_1980}. However, the radii of these vessels was much larger, $R_0=.25-1.5$ mm, with a valve spacing closer to 20 mm \citep{Macdonald_Arkill_Tabor_McHale_Winlove_2008}, so there are about 2-4 valves per wavelength. For these larger vessels, the discrete nature of the valves may be important. 

The value of $\kappa$ can be estimated by measuring the radius as a function of pressure for a nonpumping vessel and comparing to equation \eqref{eq:force_balance}. The reciprocal of the slope of the $\Delta R(\Delta P)$ curve was found to be $330 \pm 100 $ Pa/mm for excised bovine lymphatics \citep{Macdonald_Arkill_Tabor_McHale_Winlove_2008}. These results are only for a passive wall model, but for an active wall model, the Young's modulus itself can be treated as time varying \citep{Macdonald_Arkill_Tabor_McHale_Winlove_2008}. Using the values of $c$ and $\lambda$ from \cite{Ohhashi_Azuma_Sakaguchi_1980} gives a value of $\kappa > 1$. For the rat mesentery which has much smaller radius, $\kappa$ is likely small enough for compliance effects to become important. It is widely accepted in the lymphatics literature that the valves operate via the pressure drop across the valve. Further details of lymphatic valves including valve stiffness \citep{Ballard_Wolf_Nepiyushchikh_Dixon_Alexeev_2018} and hysteretic pressure response \citep{Bertram_Macaskill_Moore_2014} are neglected by our analysis. It was shown that the aspect ratio of lymphatic valves is above a critical threshold that allows for complete closure of the valves under adverse pressure conditions \citep{Ballard_Wolf_Nepiyushchikh_Dixon_Alexeev_2018}. Two works give estimates of the open valve resistance. Experimental measurements on isolated rat mesenteric lymphatics found a value of $ \mathcal{R}_v = 0.6 \times 10^6$ g/cm$^4$/s \citep{Bertram_Macaskill_Moore_2014}, while detailed modeling utilizing experimental geometric data found a value of $\mathcal{R}_v = 0.95 \times 10^6$ g/cm$^4$/s \citep{Wilson_Van_Loon_Wang_Zawieja_Moore_2015}. Using these values, and dividing by the tube resistance obtained from the rat data suggests $\Bar{r}_v = .0004 - .03$. For these larger values of $\Bar{r}_v$, the stiffness can be suppressed by a factor as large as $1+n_v \bar{r}_v=4$, which further enhances compliance effects. The fact that the elastic parameters are hard to estimate, and that $R_0$ can vary greatly within and between species, suggests the full range of $\kappa$ values should be studied, as we have in this paper, in order to understand mechanisms that could be relevant for lymphatic pumping.

\begin{table*}
  \begin{center}
\def~{\hphantom{0}}
 \begin{tabular}{c c c} 
 \hline
 Physical parameter & Measured Value & Reference  \\ 
 \hline\hline
 Viscosity $\mu$ & $8.9 \times 10^{-4}$ $\text{Pa}\cdot \text{s}$ & \cite{Macdonald_Arkill_Tabor_McHale_Winlove_2008}  \\ 
 $\Delta P/\Delta R = Eh/(1-\nu^2)R_0^2$ & $330 \pm 100$ Pa/mm & \cite{Macdonald_Arkill_Tabor_McHale_Winlove_2008}  \\
 Peristaltic period $T$ (rat) & $4.6 - 7.1$ s & \cite{Zawieja_Davis_Schuster_Hinds_Granger_1993} \\
 Peristaltic period $T$ (bovine) & $10-15$ s & \cite{Ohhashi_Azuma_Sakaguchi_1980} \\
 Rest radius $R_0$ (rat) & .03 - .06 mm & \cite{Zawieja_Davis_Schuster_Hinds_Granger_1993}  \\
  Rest radius $R_0$ (bovine) & .25 - 1.5 mm & \cite{Ohhashi_Azuma_Sakaguchi_1980}  \\
 Valve spacing $x_v$ (rat) & .6 - 1.0 mm & \cite{Zawieja_Davis_Schuster_Hinds_Granger_1993} \\ 
  Valve spacing $x_v$ (bovine) & $\approx$ 20 mm & \cite{Macdonald_Arkill_Tabor_McHale_Winlove_2008} \\ 
 Wavelength $\lambda$ (rat) & $20 - 60$ mm & calculated from \cite{Zawieja_Davis_Schuster_Hinds_Granger_1993} \\ 
  Wavelength $\lambda$ (bovine) & $40 - 75$ mm & calculated from \cite{Ohhashi_Azuma_Sakaguchi_1980} \\
 $\kappa$  & $>1$  & calculated \\ 
 $\eta_R$  & $<.5$  & \cite{Zawieja_Davis_Schuster_Hinds_Granger_1993, Davis_Scallan_Wolpers_Muthuchamy_Gashev_Zawieja_2012} \\ 
  $\bar{r}_v$  & $<.03$  & calculated from \cite{Bertram_Macaskill_Moore_2014, Wilson_Van_Loon_Wang_Zawieja_Moore_2015, Zawieja_Davis_Schuster_Hinds_Granger_1993} \\ 
 \end{tabular}
  \caption{Experimental values of physical parameters in the lymphatic system.}
  \label{tab:numerical_values}
  \end{center}
\end{table*}

The exact value of the peristaltic amplitude is unimportant, but it is reasonable to assume that the lymphatic system operates in the regime of small-amplitude peristalsis. Experimentally, one can see that the radius changes by no more than one half of its rest value \citep{Zawieja_Davis_Schuster_Hinds_Granger_1993, Davis_Scallan_Wolpers_Muthuchamy_Gashev_Zawieja_2012}. All of these considerations suggest that we should be able to at least qualitatively apply the small-amplitude valve continuum results to the lymphatic system. This work gives us some intuition for how a chain of lymphangions operates. Valves induce a mean flow at order $\eta_P$, suggesting that even small perturbations can be harnessed to drive lymph back to the circulatory system. The direction of the contraction wave is unimportant at this order, meaning no carefully coordinated, unidirectional contractions are necessary. Indeed, backward-propagating waves in lymphangions have been observed to be just as prevalent as forward-propagating waves, and the induced flows have been found to be comparable \citep{Zawieja_Davis_Schuster_Hinds_Granger_1993, McHale_Meharg_1992}.

A summary of parameters is given in table \ref{tab:parameters}. Note that these values closely resemble those reported in \cite{Wolf_Dixon_Alexeev_2021}. 

It is worth noting that for the range of parameters found for the rat, a theoretically optimum flow is achieved with tens of valves (see figure \ref{fig:nvStar}), and this is consistent with the range of $n_v$ observed. This suggests that the density of valves in the lymphatic system is large enough to rectify negative flows and not too large to unnecessarily suppress forward flows, such that $n_v \approx n_v^*$.

\subsection{Summary}
In this work, inspired by the lymphatic system, we provide some analytical and numerical results regarding fluid flow driven by peristalsis in the presence of many valves. When considered at small density, the valves produce discontinuous pressure profiles and do not support wave-like solutions typical of peristaltic pumping. Interestingly, we recover both a continuous pressure profile and wave solutions by considering the limit of an infinite density of valves. Theoretical models of peristalsis with valves have either been limited to studying only two valves treated as time-dependent boundary conditions \citep{Farina_Fusi_Fasano_Ceretani_Rosso_2016} or have relied only on numerical results \citep{Wolf_Dixon_Alexeev_2021, Ballard_Wolf_Nepiyushchikh_Dixon_Alexeev_2018}. The simplifications brought on by studying a continuum of valves have allowed us to study a range of parameters which could be relevant for explaining biological phenomena. Although the analytic results in this work were derived for the high valve density limit, the agreement between the theoretical predictions and numerical results for finite valve density were very good, even for valve densities as low as 5 valves per wavelength. This bolsters the validity of the results in more biologically relevant settings. Perhaps the most striking feature of our model is that for small-amplitude peristalsis, the mean flow grows linearly with $\eta_P$, and, in some cases, the magnitude of the flow is independent of pumping direction. This might explain the observation that both retrograde and orthograde peristaltic waves are observed in the lymphatic system with almost equal frequency \citep{Bertram_Macaskill_Moore_2014}. At large amplitudes, if peristalsis and valves are oriented the same direction, then valves do little more than increase the resistance to flow. However, if a peristaltic wave travels against the valve direction, an entirely new regime can be found where the flow is confined to a region of small radius. We also considered the effect of a finite stiffness. When $\kappa$ is infinite, radius-imposed peristalsis is recovered, with all but one valve open. Yet, even for very large finite $\kappa$, this is not observed due to a slow approach to the infinite $\kappa$ limit. When the peristaltic period is longer than the elastic relaxation time (large $\kappa$), the flow is entirely determined by the continuity equation and has a simple analytical form which can be understood as rectified diffusion that establishes a pressure gradient across the vessel. When the peristaltic period is shorter than the elastic relaxation time ($\kappa$ small), the pressure in the vessel closely resembles the applied force, so the fluid is driven by the components of peristalsis that lower the pressure downstream while the valves prevent backflow when the pressure is lower upstream. Approaching the valve continuum, the flow increases with valve density as more oscillations are rectified, but the presence of a nonzero valve resistance reduces the flow causing a decay in the mean flow proportional to $n_v^{-1}$. 

The valve continuum models an effective fluid with nonlinear properties inherited from the valves. It is tempting to associate this effective fluid with a real fluid with strange rheological properties, but the non-reciprocal response induced by valves cannot be achieved by a non-Newtonian fluid. It has been shown that for a general class of non-Newtonian fluids, it is possible to find (radius-imposed) peristaltic wave forms that induce flow against the peristalsis direction \citep{Provost_Schwarz_1994}. In our case, by construction, the flow will always be in the valve direction regardless of the peristaltic wave form, but surprisingly, the flow can even be enhanced by a backward-propagating wave, something which was not studied for non-Newtonian fluids. Effective fluids with nonlinear properties arising from flexible structures have been studied in the context of soft hair beds \citep{Alvarado_Comtet_de_Langre_Hosoi_2017}, brushes, and carpets \citep{Gopinath_Mahadevan_2011}. This is the first paper to derive the coarse-grained behavior for a system consisting of many ideal valves without appealing to lumped-parameter modeling. The methods employed to derive our two-state ideal valve continuum results can be generalized to systems containing $n$-state immersed elements provided that each ``state'' has a linear pressure-flow relationship. That is, any immersed element with a piecewise linear pressure-flow relationship can be approximated by a continuum theory with new parameters that depend on the density of the immersed elements and the resistance of each state. It is worth comparing the limit in equation \eqref{eq:approx} to that studied in \cite{Alvarado_Comtet_de_Langre_Hosoi_2017}. They considered flow over deformable hairs that are so densely packed that the flow remains confined to the tips of the hairs which may deform. In our case, the valves are still spaced far enough that the resistances of the open valves are additive. This may no longer be the case as $x_v$ becomes comparable to $R_0$. 

In order to limit the number of parameters in our model, we assumed throughout the paper that the pressure drop per wavelength was zero. The lymphatic system pumps against large adverse pressure gradients which suppress the flow \citep{Bertram_Macaskill_Moore_2014}. A nonzero pressure drop per wavelength has been taken into account in previous models of peristaltic pumping with valves \citep{Wolf_Dixon_Alexeev_2021, Farina_Fusi_Fasano_Ceretani_Rosso_2016}, but these models do not consider backward-propagating peristaltic waves. In general, the competition between the pressure gradient, peristaltic wave, and valves will determine the flow direction. 

The simplifications made in this paper will allow for the study of biologically inspired nonlinear fluidic networks without explicit dependence on the valve positions. Rather, only two numbers $\kappa$ and $\eta_P$ are needed to characterize the pumping through an edge containing many valves. This allows the lymphatic network function to be studied more easily at the whole system level and its architecture to be examined with optimality in mind, providing invaluable insights about an important, but not well understood biological flow network. 

\begin{acknowledgements}
This research was partially supported by NSF through the University of Pennsylvania Materials Research Science and Engineering Center (MRSEC) (DMR-2309043) and the Simons Foundation through Grant No. 568888.
\end{acknowledgements}

\appendix

\begin{figure*}
    \centering
    \includegraphics[width=.95\textwidth]{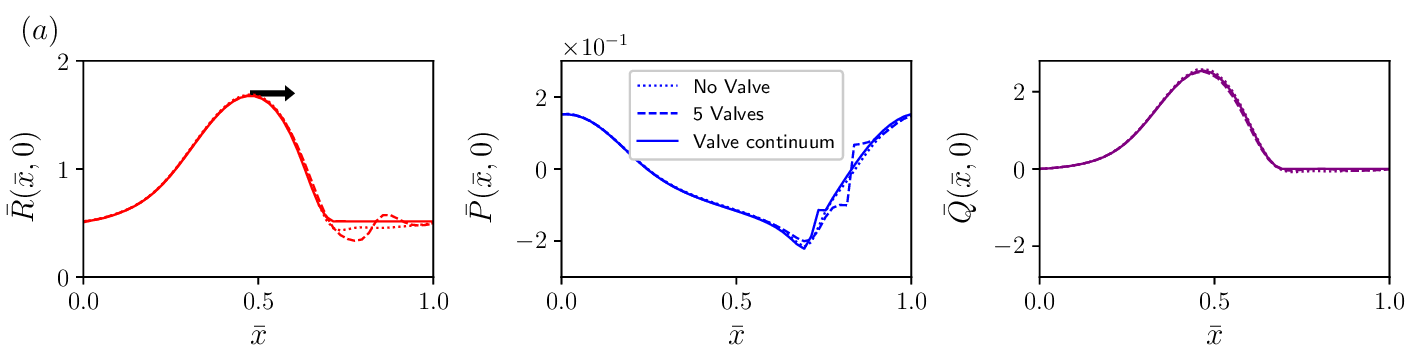}
    \includegraphics[width=.95\textwidth]{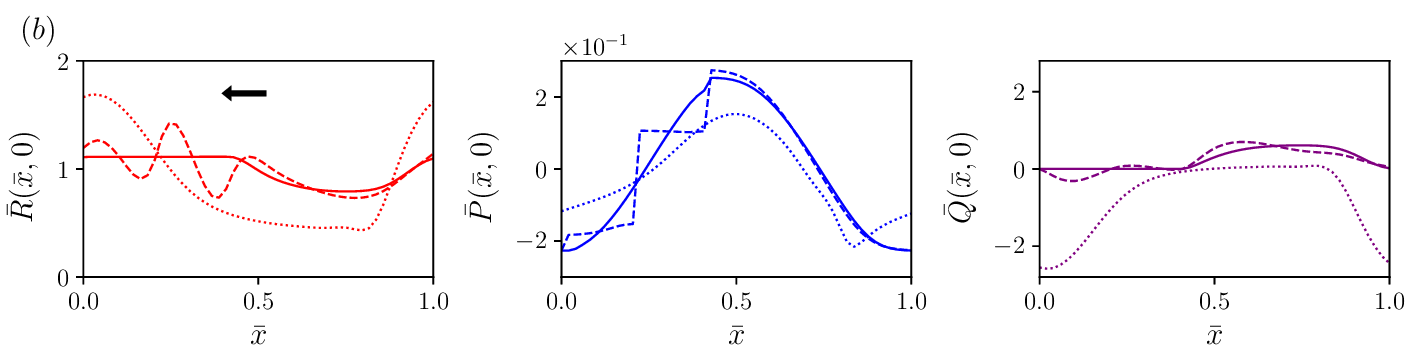}
    \caption{Results when a small bending term is introduced such that the force-balance equation takes the form \eqref{eq:PR_with_bend} with $\alpha = 10^{-7}$. Other parameters used in this simulation are $\eta_P=.25$, $\kappa=.2$, and $\bar{r}_v=0$. $(a)$ Radius, pressure, and flow induced by a forward-propagating sinusoidal peristaltic wave. The dotted line is the valveless solution, the dashed line is the solution with five equally spaced valves, and the solid line is the valve continuum result. $(b)$ Radius, pressure, and flow induced by a backward-propagating sinusoidal peristaltic wave. }
    \label{fig:bend}
\end{figure*}

\section{Bending effects}\label{appA}
Approximation \eqref{eq:approx} guarantees any force dependent on spatial derivatives of $R$ will be small. However, incorporating ideal valves into an elastic tube governed by \eqref{eq:force_balance} leads to spatial discontinuities in $R$ and cusps in $Q$ at the valve locations. In biological terms, there is no coupling between adjacent lymphangions separated by closed valves since our force balance equation \eqref{eq:force_balance} only accounts for radial forces. To couple adjacent lymphangions, we can add a small bending force so that our force-balance equation becomes
\begin{equation}
    P - P_{\text{ext}} = \frac{Eh}{(1-\nu^2)R_0} \Big[h^2 R_0^2 \frac{\partial^4 }{\partial x^4} \Big(\frac{R}{R_0}-1\Big) + \Big(\frac{R}{R_0}-1\Big)\Big]. \label{eq:PR_with_bend}
\end{equation}
 Peristalsis with bending of this form was studied in \cite{Takagi_Balmforth_2011}, and a similar term involving a second-order spatial derivative was considered in \cite{Macdonald_Arkill_Tabor_McHale_Winlove_2008}. Since $\alpha \equiv h^2R_0^2/\lambda^4 \ll 1 $, it is tempting to drop the bending terms and focus only on the stretching as we did throughout the paper. Indeed, this new term does not affect the large-scale pumping properties in the tube, justifying our use of \eqref{eq:PR}. The solutions with bend for a realistic choice of $\alpha$ (figure \ref{fig:bend}) are similar to the case without bend (figure \ref{fig:convergence}). The key difference is that the radius is now continuous and the flow is now smooth in regions of a finite number of closed valves (dashed line), but the pressure distribution remains discontinuous. The mean flow for the two cases is essentially the same. Notice that adding spatial derivatives to our force-balance equation does not affect the homogenization procedure described in section \ref{sec:valves}, so the valve continuum can easily be generalized to incorporate more complicated tube mechanics. The valve continuum still succeeds in capturing the simplified dynamics for a tube with bend, as can be seen by comparing the solid and dashed lines in figure \ref{fig:bend}.

\begin{figure*}
    \centering
    \includegraphics[width=\textwidth]{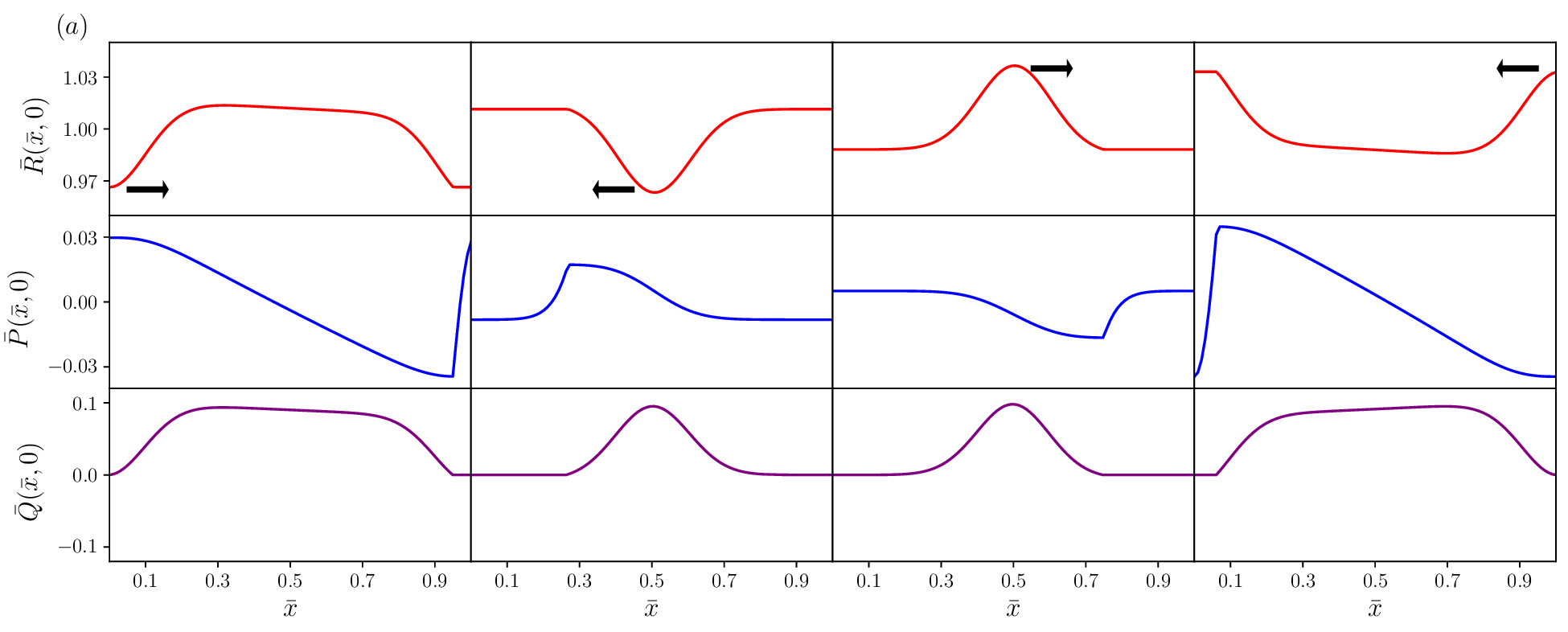}
    \includegraphics[width=\textwidth]{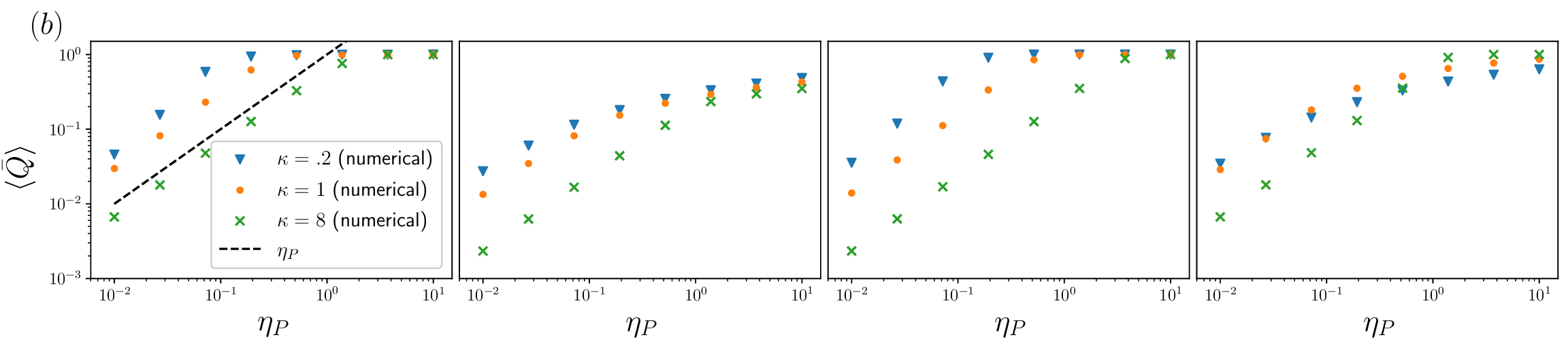}
    \caption{Summary of results for a Gaussian forcing with width parameter $l=.1$. $(a)$ Each column represents a different peristaltic gaussian wave train. In order from left to right, $f$ takes the form of a forward-propagating bell curve $f(\Bar{x},\Bar{t}) = \sum_m \exp(-(\Bar{x}-\Bar{t}-m)^2/2l^2) / \sqrt{2\pi l^2} - 1$, a backward-propagating bell curve $f(\Bar{x},\Bar{t}) = \sum_m \exp(-(\Bar{x}+\Bar{t}-m- \frac{1}{2})^2/2l^2) / \sqrt{2\pi l^2} - 1$, a forward-propagating inverted bell curve $f(\Bar{x},\Bar{t}) = 1 - \sum_m \exp(-(\Bar{x}-\Bar{t}-m - \frac{1}{2})^2/2l^2) / \sqrt{2\pi l^2}$, and a backward-propagating inverted bell curve $f(\Bar{x},\Bar{t}) = 1 - \sum_m \exp(-(\Bar{x}+\Bar{t}-m)^2/2l^2) / \sqrt{2\pi l^2}$. The radius, pressure, and flow are displayed for each. Parameters used for these simulations are $\kappa=8$, $\eta_P=0.1$, and $\Bar{r}_v=0$. Notice that when $f$ takes the form of a bell curve such that $\Bar{R}$ is an inverted bell curve, the flow is larger for a forward-propagating wave, but when $f$ takes the form of an inverted bell curve such that $\Bar{R}$ is a bell curve, the flow is larger for a backward-propagating wave. $(b)$ Mean flow as a function of amplitude for each of the cases in $(a)$.}
    \label{fig:Gauss}
\end{figure*}

\section{Gaussian forcing}
In the main text, all numerical results were given assuming $f(\xi) = \cos(2\pi \xi)$. In order to demonstrate the generality of our small-amplitude results and highlight some features absent from the sine waves, here we show the results when $f$ takes the form of a gaussian wave train. For the forward-propagating wave, $f$ takes the form
\begin{equation}
    f(\xi) = \frac{1}{\sqrt{2\pi l^2}} \sum_{m\in \mathbb{Z}} e^{-\frac{(\xi-m)^2}{2l^2}} - 1. \label{eq:gaussian}
\end{equation}
The parameter $l\ll 1$ describes the width of the gaussian. Notice that $\langle f \rangle = 0$, and $f(0)$ is a maximum so that the valve opens at $\xi=0$ for a forward-propagating wave. We also consider inverted and backward-propagating waves of a similar form. A summary of the results is given in figure \ref{fig:Gauss}. We consider four different forms of $f$ related to \eqref{eq:gaussian} by time reversal and phase shifts. The functional forms are written at the top of each column. We focus on the simple case of large $\kappa$ where the peristalsis is nearly radius imposed. By comparing the first two or last two columns in figure \ref{fig:Gauss}$a$, it is clear that the forward- and backward-propagating waves no longer produce similar flows, as was the case for sinusoidal waves in this regime. In fact, a backward-propagating inverted gaussian wave train (last column) is better at pumping than a forward-propagating inverted gaussian wave train (third column), consistent with equations \eqref{eq:QbarStiffFW} and \eqref{eq:QbarStiffBW}. There appears to be a symmetry between the forward-propagating gaussian wave and the backward-propagating inverted gaussian wave, and a symmetry between the backward-propagating gaussian wave and the forward-propagating inverted gaussian wave. This symmetry breaks down at large amplitudes as shown in figure \ref{fig:Gauss}$b$. For larger amplitudes, both forward-propagating waves approach $\langle \bar{Q} \rangle =1$ faster than either backward-propagating wave. 

As this example demonstrates, the amplitude is not the only important feature of the peristaltic wave. Certain shapes are more effective at pumping in certain directions than others. This is reminiscent of the findings in \cite{Provost_Schwarz_1994} who consider how to engineer peristaltic waves to optimize retrograde flow due to non-Newtonian fluid properties as opposed to valves. 
\\
\section{Table of parameters}
See table \ref{tab:parameters} for a summary of parameters used in the paper. 

\begin{table*}
  \begin{center}
\def~{\hphantom{0}}

\begin{tabular}{c  c} 
 \hline
 Symbol & Description \\
 \hline\hline
 $R_0$ & Radius of the pipe when no external forces are present \\ 
 $x_v^i$ & Position of the $i^{\text{th}}$ valve \\ 
  $x_v$ & Mean valve spacing \\ 
   $\lambda$ & Wavelength of peristalsis \\ 
   $\epsilon$ & $x_v/\lambda$ \\ 
    $n_v$ & Number of valves per wavelength = $\epsilon^{-1}$ \\ 
     $\bar{r}_v$ & Valve resistance parameter as defined in equation \eqref{eq:rv} \\ 
    $\eta_P$ & Strength of applied peristaltic forces as defined in equation \eqref{eq:parameters}\\ 
    $\eta_R$ & Strength of radial deformation induced by peristalsis as defined in equation \eqref{eq:parameters}\\ 
     $T$ & Period of peristaltic wave \\ 
     $c$ & Peristaltic wave speed $=\lambda/T$\\ 
    $\kappa$ & Dimensionless stiffness parameter as defined in \eqref{eq:parameters}\\ 
    $\alpha$ &  Strength of bend = $h^2 R_0^2/\lambda^2$\\     
       
    $P_a$ &  Characteristic amplitude of peristalsis \\  
     $\bar{x}$ &  Dimensionless axial length used to study fluid dynamics at large length scales $x/\lambda$ \\  
     $\bar{y}$ &  Dimensionless axial length used to study the region between closely spaced valves $x/\epsilon$ \\  
     \\[-1em]
     $\bar{t}$ &  Dimensionless time $t/T$ \\      
     $\xi$ &  $\bar{x}-\bar{t}$ for a forward-propagating wave, or $\bar{x}+\bar{t}$ for a backward-propagating wave  \\  
    $f(x,t)$ &  Dimensionless peristaltic force imposed on the tube\\ 
    $P$, $Q$, $R$ &  Fluid pressure, volumetric flow rate, and tube radius \\ 
    \\[-1em]
    $\bar{P}$, $\bar{Q}$, $\bar{R}$ &  Dimensionless pressure, flow, and radius, as defined by equations \eqref{eq:PQR_nondim} \\ 
    $\langle Q \rangle $ &  Angle brackets denote an average over one period  \\     
    $P_i$ &  Subscript $i$ denotes the $i$th term in an expansion in powers of a small parameter ($\epsilon$ or $\eta_P$)\\
    \\[-1em]
     $\tilde{\xi}$ &  Coordinate at which the valve opens. \\     
      $n_v^*$ &  The value of $n_v$ which maximizes the mean flow. \\
      $\Delta P_v^i$ & Pressure drop across valve $i$ (upstream pressure minus downstream pressure)\\
      $Q^{\text{nv}}$ & Superscript ``nv" denotes that the tube contains no valves
 
\end{tabular}
  \caption{Parameters used in the paper.}
  \label{tab:parameters}
  \end{center}
\end{table*}

\bibliography{refs}

\begin{thebibliography}{30}%
\makeatletter
\providecommand \@ifxundefined [1]{%
 \@ifx{#1\undefined}
}%
\providecommand \@ifnum [1]{%
 \ifnum #1\expandafter \@firstoftwo
 \else \expandafter \@secondoftwo
 \fi
}%
\providecommand \@ifx [1]{%
 \ifx #1\expandafter \@firstoftwo
 \else \expandafter \@secondoftwo
 \fi
}%
\providecommand \natexlab [1]{#1}%
\providecommand \enquote  [1]{``#1''}%
\providecommand \bibnamefont  [1]{#1}%
\providecommand \bibfnamefont [1]{#1}%
\providecommand \citenamefont [1]{#1}%
\providecommand \href@noop [0]{\@secondoftwo}%
\providecommand \href [0]{\begingroup \@sanitize@url \@href}%
\providecommand \@href[1]{\@@startlink{#1}\@@href}%
\providecommand \@@href[1]{\endgroup#1\@@endlink}%
\providecommand \@sanitize@url [0]{\catcode `\\12\catcode `\$12\catcode `\&12\catcode `\#12\catcode `\^12\catcode `\_12\catcode `\%12\relax}%
\providecommand \@@startlink[1]{}%
\providecommand \@@endlink[0]{}%
\providecommand \url  [0]{\begingroup\@sanitize@url \@url }%
\providecommand \@url [1]{\endgroup\@href {#1}{\urlprefix }}%
\providecommand \urlprefix  [0]{URL }%
\providecommand \Eprint [0]{\href }%
\providecommand \doibase [0]{https://doi.org/}%
\providecommand \selectlanguage [0]{\@gobble}%
\providecommand \bibinfo  [0]{\@secondoftwo}%
\providecommand \bibfield  [0]{\@secondoftwo}%
\providecommand \translation [1]{[#1]}%
\providecommand \BibitemOpen [0]{}%
\providecommand \bibitemStop [0]{}%
\providecommand \bibitemNoStop [0]{.\EOS\space}%
\providecommand \EOS [0]{\spacefactor3000\relax}%
\providecommand \BibitemShut  [1]{\csname bibitem#1\endcsname}%
\let\auto@bib@innerbib\@empty
\bibitem [{\citenamefont {Brasseur}(1987)}]{Brasseur_1987}%
  \BibitemOpen
  \bibfield  {author} {\bibinfo {author} {\bibfnamefont {J.~G.}\ \bibnamefont {Brasseur}},\ }\href {https://doi.org/10.1007/BF02406976} {\bibfield  {journal} {\bibinfo  {journal} {Dysphagia}\ }\textbf {\bibinfo {volume} {2}},\ \bibinfo {pages} {32–39} (\bibinfo {year} {1987})}\BibitemShut {NoStop}%
\bibitem [{\citenamefont {Carew}\ and\ \citenamefont {Pedley}(1997)}]{Carew_Pedley_1997}%
  \BibitemOpen
  \bibfield  {author} {\bibinfo {author} {\bibfnamefont {E.~O.}\ \bibnamefont {Carew}}\ and\ \bibinfo {author} {\bibfnamefont {T.~J.}\ \bibnamefont {Pedley}},\ }\href {https://doi.org/10.1115/1.2796066} {\bibfield  {journal} {\bibinfo  {journal} {Journal of Biomechanical Engineering}\ }\textbf {\bibinfo {volume} {119}},\ \bibinfo {pages} {66–76} (\bibinfo {year} {1997})}\BibitemShut {NoStop}%
\bibitem [{\citenamefont {Moore}\ and\ \citenamefont {Bertram}(2018)}]{Moore_Bertram_2018}%
  \BibitemOpen
  \bibfield  {author} {\bibinfo {author} {\bibfnamefont {J.~E.}\ \bibnamefont {Moore}}\ and\ \bibinfo {author} {\bibfnamefont {C.~D.}\ \bibnamefont {Bertram}},\ }\href {https://doi.org/10.1146/annurev-fluid-122316-045259} {\bibfield  {journal} {\bibinfo  {journal} {Annual review of fluid mechanics}\ }\textbf {\bibinfo {volume} {50}},\ \bibinfo {pages} {459–482} (\bibinfo {year} {2018})}\BibitemShut {NoStop}%
\bibitem [{\citenamefont {Carr}\ \emph {et~al.}(2021)\citenamefont {Carr}, \citenamefont {Thomas}, \citenamefont {Liu},\ and\ \citenamefont {Shang}}]{carr_thomas_liu_shang_2021}%
  \BibitemOpen
  \bibfield  {author} {\bibinfo {author} {\bibfnamefont {J.~B.}\ \bibnamefont {Carr}}, \bibinfo {author} {\bibfnamefont {J.~H.}\ \bibnamefont {Thomas}}, \bibinfo {author} {\bibfnamefont {J.}~\bibnamefont {Liu}},\ and\ \bibinfo {author} {\bibfnamefont {J.~K.}\ \bibnamefont {Shang}},\ }\href {https://doi.org/10.1017/jfm.2021.277} {\bibfield  {journal} {\bibinfo  {journal} {Journal of Fluid Mechanics}\ }\textbf {\bibinfo {volume} {917}},\ \bibinfo {pages} {A10} (\bibinfo {year} {2021})}\BibitemShut {NoStop}%
\bibitem [{\citenamefont {Mestre}\ \emph {et~al.}(2018)\citenamefont {Mestre}, \citenamefont {Tithof}, \citenamefont {Du}, \citenamefont {Song}, \citenamefont {Peng}, \citenamefont {Sweeney}, \citenamefont {Olveda}, \citenamefont {Thomas}, \citenamefont {Nedergaard},\ and\ \citenamefont {Kelley}}]{Mestre2018}%
  \BibitemOpen
  \bibfield  {author} {\bibinfo {author} {\bibfnamefont {H.}~\bibnamefont {Mestre}}, \bibinfo {author} {\bibfnamefont {J.}~\bibnamefont {Tithof}}, \bibinfo {author} {\bibfnamefont {T.}~\bibnamefont {Du}}, \bibinfo {author} {\bibfnamefont {W.}~\bibnamefont {Song}}, \bibinfo {author} {\bibfnamefont {W.}~\bibnamefont {Peng}}, \bibinfo {author} {\bibfnamefont {A.~M.}\ \bibnamefont {Sweeney}}, \bibinfo {author} {\bibfnamefont {G.}~\bibnamefont {Olveda}}, \bibinfo {author} {\bibfnamefont {J.~H.}\ \bibnamefont {Thomas}}, \bibinfo {author} {\bibfnamefont {M.}~\bibnamefont {Nedergaard}},\ and\ \bibinfo {author} {\bibfnamefont {D.~H.}\ \bibnamefont {Kelley}},\ }\bibfield  {journal} {\bibinfo  {journal} {Nature Communications}\ }\textbf {\bibinfo {volume} {9}},\ \href {https://doi.org/10.1038/s41467-018-07318-3} {10.1038/s41467-018-07318-3} (\bibinfo {year} {2018})\BibitemShut {NoStop}%
\bibitem [{\citenamefont {Burns}\ and\ \citenamefont {Parkes}(1967)}]{Burns_Parkes_1967}%
  \BibitemOpen
  \bibfield  {author} {\bibinfo {author} {\bibfnamefont {J.~C.}\ \bibnamefont {Burns}}\ and\ \bibinfo {author} {\bibfnamefont {T.}~\bibnamefont {Parkes}},\ }\href {https://doi.org/10.1017/S0022112067001156} {\bibfield  {journal} {\bibinfo  {journal} {Journal of Fluid Mechanics}\ }\textbf {\bibinfo {volume} {29}},\ \bibinfo {pages} {731–743} (\bibinfo {year} {1967})}\BibitemShut {NoStop}%
\bibitem [{\citenamefont {Shapiro}\ \emph {et~al.}(1969)\citenamefont {Shapiro}, \citenamefont {Jaffrin},\ and\ \citenamefont {Weinberg}}]{Shapiro_Jaffrin_Weinberg_1969}%
  \BibitemOpen
  \bibfield  {author} {\bibinfo {author} {\bibfnamefont {A.~H.}\ \bibnamefont {Shapiro}}, \bibinfo {author} {\bibfnamefont {M.~Y.}\ \bibnamefont {Jaffrin}},\ and\ \bibinfo {author} {\bibfnamefont {S.~L.}\ \bibnamefont {Weinberg}},\ }\href {https://doi.org/10.1017/S0022112069000899} {\bibfield  {journal} {\bibinfo  {journal} {Journal of Fluid Mechanics}\ }\textbf {\bibinfo {volume} {37}},\ \bibinfo {pages} {799–825} (\bibinfo {year} {1969})}\BibitemShut {NoStop}%
\bibitem [{\citenamefont {Farina}\ \emph {et~al.}(2016)\citenamefont {Farina}, \citenamefont {Fusi}, \citenamefont {Fasano}, \citenamefont {Ceretani},\ and\ \citenamefont {Rosso}}]{Farina_Fusi_Fasano_Ceretani_Rosso_2016}%
  \BibitemOpen
  \bibfield  {author} {\bibinfo {author} {\bibfnamefont {A.}~\bibnamefont {Farina}}, \bibinfo {author} {\bibfnamefont {L.}~\bibnamefont {Fusi}}, \bibinfo {author} {\bibfnamefont {A.}~\bibnamefont {Fasano}}, \bibinfo {author} {\bibfnamefont {A.}~\bibnamefont {Ceretani}},\ and\ \bibinfo {author} {\bibfnamefont {F.}~\bibnamefont {Rosso}},\ }\href {https://doi.org/10.1016/j.ijengsci.2016.07.002} {\bibfield  {journal} {\bibinfo  {journal} {International Journal of Engineering Science}\ }\textbf {\bibinfo {volume} {107}},\ \bibinfo {pages} {1–12} (\bibinfo {year} {2016})}\BibitemShut {NoStop}%
\bibitem [{\citenamefont {Margaris}\ and\ \citenamefont {Black}(2012)}]{Margaris_Black_2012}%
  \BibitemOpen
  \bibfield  {author} {\bibinfo {author} {\bibfnamefont {K.~N.}\ \bibnamefont {Margaris}}\ and\ \bibinfo {author} {\bibfnamefont {R.~A.}\ \bibnamefont {Black}},\ }\href {https://doi.org/10.1098/rsif.2011.0751} {\bibfield  {journal} {\bibinfo  {journal} {Journal of The Royal Society Interface}\ }\textbf {\bibinfo {volume} {9}},\ \bibinfo {pages} {601–612} (\bibinfo {year} {2012})}\BibitemShut {NoStop}%
\bibitem [{\citenamefont {McHale}\ and\ \citenamefont {Meharg}(1992)}]{McHale_Meharg_1992}%
  \BibitemOpen
  \bibfield  {author} {\bibinfo {author} {\bibfnamefont {N.~G.}\ \bibnamefont {McHale}}\ and\ \bibinfo {author} {\bibfnamefont {M.~K.}\ \bibnamefont {Meharg}},\ }\href {https://doi.org/10.1113/jphysiol.1992.sp019139} {\bibfield  {journal} {\bibinfo  {journal} {The Journal of Physiology}\ }\textbf {\bibinfo {volume} {450}},\ \bibinfo {pages} {503–512} (\bibinfo {year} {1992})}\BibitemShut {NoStop}%
\bibitem [{\citenamefont {Zawieja}\ \emph {et~al.}(1993)\citenamefont {Zawieja}, \citenamefont {Davis}, \citenamefont {Schuster}, \citenamefont {Hinds},\ and\ \citenamefont {Granger}}]{Zawieja_Davis_Schuster_Hinds_Granger_1993}%
  \BibitemOpen
  \bibfield  {author} {\bibinfo {author} {\bibfnamefont {D.~C.}\ \bibnamefont {Zawieja}}, \bibinfo {author} {\bibfnamefont {K.~L.}\ \bibnamefont {Davis}}, \bibinfo {author} {\bibfnamefont {R.}~\bibnamefont {Schuster}}, \bibinfo {author} {\bibfnamefont {W.~M.}\ \bibnamefont {Hinds}},\ and\ \bibinfo {author} {\bibfnamefont {H.~J.}\ \bibnamefont {Granger}},\ }\href {https://doi.org/10.1152/ajpheart.1993.264.4.H1283} {\bibfield  {journal} {\bibinfo  {journal} {American Journal of Physiology-Heart and Circulatory Physiology}\ }\textbf {\bibinfo {volume} {264}},\ \bibinfo {pages} {H1283–H1291} (\bibinfo {year} {1993})}\BibitemShut {NoStop}%
\bibitem [{\citenamefont {Ballard}\ \emph {et~al.}(2018)\citenamefont {Ballard}, \citenamefont {Wolf}, \citenamefont {Nepiyushchikh}, \citenamefont {Dixon},\ and\ \citenamefont {Alexeev}}]{Ballard_Wolf_Nepiyushchikh_Dixon_Alexeev_2018}%
  \BibitemOpen
  \bibfield  {author} {\bibinfo {author} {\bibfnamefont {M.}~\bibnamefont {Ballard}}, \bibinfo {author} {\bibfnamefont {K.~T.}\ \bibnamefont {Wolf}}, \bibinfo {author} {\bibfnamefont {Z.}~\bibnamefont {Nepiyushchikh}}, \bibinfo {author} {\bibfnamefont {J.~B.}\ \bibnamefont {Dixon}},\ and\ \bibinfo {author} {\bibfnamefont {A.}~\bibnamefont {Alexeev}},\ }\href {https://doi.org/10.1007/s10237-018-1030-y} {\bibfield  {journal} {\bibinfo  {journal} {Biomechanics and Modeling in Mechanobiology}\ }\textbf {\bibinfo {volume} {17}},\ \bibinfo {pages} {1343–1356} (\bibinfo {year} {2018})}\BibitemShut {NoStop}%
\bibitem [{\citenamefont {Wolf}\ \emph {et~al.}(2021)\citenamefont {Wolf}, \citenamefont {Dixon},\ and\ \citenamefont {Alexeev}}]{Wolf_Dixon_Alexeev_2021}%
  \BibitemOpen
  \bibfield  {author} {\bibinfo {author} {\bibfnamefont {K.~T.}\ \bibnamefont {Wolf}}, \bibinfo {author} {\bibfnamefont {J.~B.}\ \bibnamefont {Dixon}},\ and\ \bibinfo {author} {\bibfnamefont {A.}~\bibnamefont {Alexeev}},\ }\href {https://doi.org/10.1017/jfm.2021.302} {\bibfield  {journal} {\bibinfo  {journal} {Journal of Fluid Mechanics}\ }\textbf {\bibinfo {volume} {918}},\ \bibinfo {pages} {A28} (\bibinfo {year} {2021})}\BibitemShut {NoStop}%
\bibitem [{\citenamefont {Wolf}\ \emph {et~al.}(2023)\citenamefont {Wolf}, \citenamefont {Poorghani}, \citenamefont {Dixon},\ and\ \citenamefont {Alexeev}}]{Wolf_Poorghani_Dixon_Alexeev_2023}%
  \BibitemOpen
  \bibfield  {author} {\bibinfo {author} {\bibfnamefont {K.~T.}\ \bibnamefont {Wolf}}, \bibinfo {author} {\bibfnamefont {A.}~\bibnamefont {Poorghani}}, \bibinfo {author} {\bibfnamefont {J.~B.}\ \bibnamefont {Dixon}},\ and\ \bibinfo {author} {\bibfnamefont {A.}~\bibnamefont {Alexeev}},\ }\href {https://doi.org/10.1088/1748-3190/acbe85} {\bibfield  {journal} {\bibinfo  {journal} {Bioinspiration and Biomimetics}\ }\textbf {\bibinfo {volume} {18}},\ \bibinfo {pages} {035002} (\bibinfo {year} {2023})}\BibitemShut {NoStop}%
\bibitem [{\citenamefont {Park}\ \emph {et~al.}(2018)\citenamefont {Park}, \citenamefont {Tixier}, \citenamefont {Christensen}, \citenamefont {Arnbjerg-Nielsen}, \citenamefont {Zwieniecki},\ and\ \citenamefont {Jensen}}]{Park2018}%
  \BibitemOpen
  \bibfield  {author} {\bibinfo {author} {\bibfnamefont {K.}~\bibnamefont {Park}}, \bibinfo {author} {\bibfnamefont {A.}~\bibnamefont {Tixier}}, \bibinfo {author} {\bibfnamefont {A.~H.}\ \bibnamefont {Christensen}}, \bibinfo {author} {\bibfnamefont {S.~F.}\ \bibnamefont {Arnbjerg-Nielsen}}, \bibinfo {author} {\bibfnamefont {M.~A.}\ \bibnamefont {Zwieniecki}},\ and\ \bibinfo {author} {\bibfnamefont {K.~H.}\ \bibnamefont {Jensen}},\ }\href {https://doi.org/10.1017/jfm.2017.805} {\bibfield  {journal} {\bibinfo  {journal} {Journal of Fluid Mechanics}\ }\textbf {\bibinfo {volume} {836}},\ \bibinfo {pages} {R3} (\bibinfo {year} {2018})},\ \Eprint {https://arxiv.org/abs/1708.06968} {arXiv:1708.06968} \BibitemShut {NoStop}%
\bibitem [{\citenamefont {Brandenbourger}\ \emph {et~al.}(2020)\citenamefont {Brandenbourger}, \citenamefont {Dangremont}, \citenamefont {Sprik},\ and\ \citenamefont {Coulais}}]{Brandenbourger2020}%
  \BibitemOpen
  \bibfield  {author} {\bibinfo {author} {\bibfnamefont {M.}~\bibnamefont {Brandenbourger}}, \bibinfo {author} {\bibfnamefont {A.}~\bibnamefont {Dangremont}}, \bibinfo {author} {\bibfnamefont {R.}~\bibnamefont {Sprik}},\ and\ \bibinfo {author} {\bibfnamefont {C.}~\bibnamefont {Coulais}},\ }\href {https://doi.org/10.1103/PhysRevFluids.5.084102} {\bibfield  {journal} {\bibinfo  {journal} {Physical Review Fluids}\ }\textbf {\bibinfo {volume} {5}},\ \bibinfo {pages} {1} (\bibinfo {year} {2020})}\BibitemShut {NoStop}%
\bibitem [{\citenamefont {Takagi}\ and\ \citenamefont {Balmforth}(2011)}]{Takagi_Balmforth_2011}%
  \BibitemOpen
  \bibfield  {author} {\bibinfo {author} {\bibfnamefont {D.}~\bibnamefont {Takagi}}\ and\ \bibinfo {author} {\bibfnamefont {N.}~\bibnamefont {Balmforth}},\ }\href {https://doi.org/10.1017/S0022112010005914} {\bibfield  {journal} {\bibinfo  {journal} {Journal of Fluid Mechanics}\ }\textbf {\bibinfo {volume} {672}},\ \bibinfo {pages} {196 } (\bibinfo {year} {2011})}\BibitemShut {NoStop}%
\bibitem [{\citenamefont {Elbaz}\ and\ \citenamefont {Gat}(2014)}]{Elbaz_Gat_2014}%
  \BibitemOpen
  \bibfield  {author} {\bibinfo {author} {\bibfnamefont {S.~B.}\ \bibnamefont {Elbaz}}\ and\ \bibinfo {author} {\bibfnamefont {A.~D.}\ \bibnamefont {Gat}},\ }\href {https://doi.org/10.1017/jfm.2014.527} {\bibfield  {journal} {\bibinfo  {journal} {Journal of Fluid Mechanics}\ }\textbf {\bibinfo {volume} {758}},\ \bibinfo {pages} {221–237} (\bibinfo {year} {2014})}\BibitemShut {NoStop}%
\bibitem [{\citenamefont {Timoshenko}\ and\ \citenamefont {Woinowsky-Krieger}(1959)}]{Timoshenko_1959}%
  \BibitemOpen
  \bibfield  {author} {\bibinfo {author} {\bibfnamefont {S.}~\bibnamefont {Timoshenko}}\ and\ \bibinfo {author} {\bibfnamefont {S.}~\bibnamefont {Woinowsky-Krieger}},\ }\href@noop {} {\emph {\bibinfo {title} {Theory of Plates and Shells}}}\ (\bibinfo  {publisher} {McGraw-Hill},\ \bibinfo {year} {1959})\BibitemShut {NoStop}%
\bibitem [{\citenamefont {Macdonald}\ \emph {et~al.}(2008)\citenamefont {Macdonald}, \citenamefont {Arkill}, \citenamefont {Tabor}, \citenamefont {McHale},\ and\ \citenamefont {Winlove}}]{Macdonald_Arkill_Tabor_McHale_Winlove_2008}%
  \BibitemOpen
  \bibfield  {author} {\bibinfo {author} {\bibfnamefont {A.~J.}\ \bibnamefont {Macdonald}}, \bibinfo {author} {\bibfnamefont {K.~P.}\ \bibnamefont {Arkill}}, \bibinfo {author} {\bibfnamefont {G.~R.}\ \bibnamefont {Tabor}}, \bibinfo {author} {\bibfnamefont {N.~G.}\ \bibnamefont {McHale}},\ and\ \bibinfo {author} {\bibfnamefont {C.~P.}\ \bibnamefont {Winlove}},\ }\href {https://doi.org/10.1152/ajpheart.00004.2008} {\bibfield  {journal} {\bibinfo  {journal} {American Journal of Physiology-Heart and Circulatory Physiology}\ }\textbf {\bibinfo {volume} {295}},\ \bibinfo {pages} {H305–H313} (\bibinfo {year} {2008})}\BibitemShut {NoStop}%
\bibitem [{\citenamefont {Provost}\ and\ \citenamefont {Schwarz}(1994)}]{Provost_Schwarz_1994}%
  \BibitemOpen
  \bibfield  {author} {\bibinfo {author} {\bibfnamefont {A.~M.}\ \bibnamefont {Provost}}\ and\ \bibinfo {author} {\bibfnamefont {W.~H.}\ \bibnamefont {Schwarz}},\ }\href {https://doi.org/10.1017/S0022112094003873} {\bibfield  {journal} {\bibinfo  {journal} {Journal of Fluid Mechanics}\ }\textbf {\bibinfo {volume} {279}},\ \bibinfo {pages} {177–195} (\bibinfo {year} {1994})}\BibitemShut {NoStop}%
\bibitem [{\citenamefont {Holmes}(2013)}]{Holmes_2013}%
  \BibitemOpen
  \bibfield  {author} {\bibinfo {author} {\bibfnamefont {M.~H.}\ \bibnamefont {Holmes}},\ }\href@noop {} {\emph {\bibinfo {title} {Introduction to perturbation methods}}},\ \bibinfo {edition} {2nd}\ ed.\ (\bibinfo  {publisher} {Springer},\ \bibinfo {year} {2013})\BibitemShut {NoStop}%
\bibitem [{\citenamefont {Venugopal}\ \emph {et~al.}(2009)\citenamefont {Venugopal}, \citenamefont {Quick}, \citenamefont {Laine},\ and\ \citenamefont {Stewart}}]{Venugopal2009a}%
  \BibitemOpen
  \bibfield  {author} {\bibinfo {author} {\bibfnamefont {A.~M.}\ \bibnamefont {Venugopal}}, \bibinfo {author} {\bibfnamefont {C.~M.}\ \bibnamefont {Quick}}, \bibinfo {author} {\bibfnamefont {G.~A.}\ \bibnamefont {Laine}},\ and\ \bibinfo {author} {\bibfnamefont {R.~H.}\ \bibnamefont {Stewart}},\ }\href {https://doi.org/10.1152/ajpheart.00360.2008} {\bibfield  {journal} {\bibinfo  {journal} {American Journal of Physiology - Heart and Circulatory Physiology}\ }\textbf {\bibinfo {volume} {296}},\ \bibinfo {pages} {303} (\bibinfo {year} {2009})}\BibitemShut {NoStop}%
\bibitem [{\citenamefont {Dixon}\ \emph {et~al.}(2006)\citenamefont {Dixon}, \citenamefont {Greiner}, \citenamefont {Gashev}, \citenamefont {Cote}, \citenamefont {Moore},\ and\ \citenamefont {Zawieja}}]{Dixon_Greiner_Gashev_Cote_Moore_Zawieja_2006}%
  \BibitemOpen
  \bibfield  {author} {\bibinfo {author} {\bibfnamefont {J.~B.}\ \bibnamefont {Dixon}}, \bibinfo {author} {\bibfnamefont {S.~T.}\ \bibnamefont {Greiner}}, \bibinfo {author} {\bibfnamefont {A.~A.}\ \bibnamefont {Gashev}}, \bibinfo {author} {\bibfnamefont {G.~L.}\ \bibnamefont {Cote}}, \bibinfo {author} {\bibfnamefont {J.~E.}\ \bibnamefont {Moore}},\ and\ \bibinfo {author} {\bibfnamefont {D.~C.}\ \bibnamefont {Zawieja}},\ }\href {https://doi.org/10.1080/10739680600893909} {\bibfield  {journal} {\bibinfo  {journal} {Microcirculation}\ }\textbf {\bibinfo {volume} {13}},\ \bibinfo {pages} {597–610} (\bibinfo {year} {2006})}\BibitemShut {NoStop}%
\bibitem [{\citenamefont {Ohhashi}\ \emph {et~al.}(1980)\citenamefont {Ohhashi}, \citenamefont {Azuma},\ and\ \citenamefont {Sakaguchi}}]{Ohhashi_Azuma_Sakaguchi_1980}%
  \BibitemOpen
  \bibfield  {author} {\bibinfo {author} {\bibfnamefont {T.}~\bibnamefont {Ohhashi}}, \bibinfo {author} {\bibfnamefont {T.}~\bibnamefont {Azuma}},\ and\ \bibinfo {author} {\bibfnamefont {M.}~\bibnamefont {Sakaguchi}},\ }\href {https://doi.org/10.1152/ajpheart.1980.239.1.H88} {\bibfield  {journal} {\bibinfo  {journal} {American Journal of Physiology-Heart and Circulatory Physiology}\ }\textbf {\bibinfo {volume} {239}},\ \bibinfo {pages} {H88–H95} (\bibinfo {year} {1980})}\BibitemShut {NoStop}%
\bibitem [{\citenamefont {Bertram}\ \emph {et~al.}(2014)\citenamefont {Bertram}, \citenamefont {Macaskill},\ and\ \citenamefont {Moore}}]{Bertram_Macaskill_Moore_2014}%
  \BibitemOpen
  \bibfield  {author} {\bibinfo {author} {\bibfnamefont {C.~D.}\ \bibnamefont {Bertram}}, \bibinfo {author} {\bibfnamefont {C.}~\bibnamefont {Macaskill}},\ and\ \bibinfo {author} {\bibfnamefont {J.~E.}\ \bibnamefont {Moore}},\ }\href {https://doi.org/10.1080/10255842.2012.753066} {\bibfield  {journal} {\bibinfo  {journal} {Computer Methods in Biomechanics and Biomedical Engineering}\ }\textbf {\bibinfo {volume} {17}},\ \bibinfo {pages} {1519–1534} (\bibinfo {year} {2014})}\BibitemShut {NoStop}%
\bibitem [{\citenamefont {Wilson}\ \emph {et~al.}(2015)\citenamefont {Wilson}, \citenamefont {Van~Loon}, \citenamefont {Wang}, \citenamefont {Zawieja},\ and\ \citenamefont {Moore}}]{Wilson_Van_Loon_Wang_Zawieja_Moore_2015}%
  \BibitemOpen
  \bibfield  {author} {\bibinfo {author} {\bibfnamefont {J.~T.}\ \bibnamefont {Wilson}}, \bibinfo {author} {\bibfnamefont {R.}~\bibnamefont {Van~Loon}}, \bibinfo {author} {\bibfnamefont {W.}~\bibnamefont {Wang}}, \bibinfo {author} {\bibfnamefont {D.~C.}\ \bibnamefont {Zawieja}},\ and\ \bibinfo {author} {\bibfnamefont {J.~E.}\ \bibnamefont {Moore}},\ }\href {https://doi.org/10.1016/j.jbiomech.2015.07.045} {\bibfield  {journal} {\bibinfo  {journal} {Journal of Biomechanics}\ }\textbf {\bibinfo {volume} {48}},\ \bibinfo {pages} {3584–3590} (\bibinfo {year} {2015})}\BibitemShut {NoStop}%
\bibitem [{\citenamefont {Davis}\ \emph {et~al.}(2012)\citenamefont {Davis}, \citenamefont {Scallan}, \citenamefont {Wolpers}, \citenamefont {Muthuchamy}, \citenamefont {Gashev},\ and\ \citenamefont {Zawieja}}]{Davis_Scallan_Wolpers_Muthuchamy_Gashev_Zawieja_2012}%
  \BibitemOpen
  \bibfield  {author} {\bibinfo {author} {\bibfnamefont {M.~J.}\ \bibnamefont {Davis}}, \bibinfo {author} {\bibfnamefont {J.~P.}\ \bibnamefont {Scallan}}, \bibinfo {author} {\bibfnamefont {J.~H.}\ \bibnamefont {Wolpers}}, \bibinfo {author} {\bibfnamefont {M.}~\bibnamefont {Muthuchamy}}, \bibinfo {author} {\bibfnamefont {A.~A.}\ \bibnamefont {Gashev}},\ and\ \bibinfo {author} {\bibfnamefont {D.~C.}\ \bibnamefont {Zawieja}},\ }\href {https://doi.org/10.1152/ajpheart.01097.2011} {\bibfield  {journal} {\bibinfo  {journal} {American Journal of Physiology-Heart and Circulatory Physiology}\ }\textbf {\bibinfo {volume} {303}},\ \bibinfo {pages} {H795–H808} (\bibinfo {year} {2012})}\BibitemShut {NoStop}%
\bibitem [{\citenamefont {Alvarado}\ \emph {et~al.}(2017)\citenamefont {Alvarado}, \citenamefont {Comtet}, \citenamefont {de~Langre},\ and\ \citenamefont {Hosoi}}]{Alvarado_Comtet_de_Langre_Hosoi_2017}%
  \BibitemOpen
  \bibfield  {author} {\bibinfo {author} {\bibfnamefont {J.}~\bibnamefont {Alvarado}}, \bibinfo {author} {\bibfnamefont {J.}~\bibnamefont {Comtet}}, \bibinfo {author} {\bibfnamefont {E.}~\bibnamefont {de~Langre}},\ and\ \bibinfo {author} {\bibfnamefont {A.~E.}\ \bibnamefont {Hosoi}},\ }\href {https://doi.org/10.1038/nphys4225} {\bibfield  {journal} {\bibinfo  {journal} {Nature Physics}\ }\textbf {\bibinfo {volume} {13}},\ \bibinfo {pages} {1014–1019} (\bibinfo {year} {2017})}\BibitemShut {NoStop}%
\bibitem [{\citenamefont {Gopinath}\ and\ \citenamefont {Mahadevan}(2011)}]{Gopinath_Mahadevan_2011}%
  \BibitemOpen
  \bibfield  {author} {\bibinfo {author} {\bibfnamefont {A.}~\bibnamefont {Gopinath}}\ and\ \bibinfo {author} {\bibfnamefont {L.}~\bibnamefont {Mahadevan}},\ }\href {https://doi.org/10.1098/rspa.2010.0228} {\bibfield  {journal} {\bibinfo  {journal} {Proceedings of the Royal Society A: Mathematical, Physical and Engineering Sciences}\ }\textbf {\bibinfo {volume} {467}},\ \bibinfo {pages} {1665–1685} (\bibinfo {year} {2011})}\BibitemShut {NoStop}%
\end{thebibliography}%

\end{document}